\documentclass[prx,floatfix,showpacs,superscriptaddress,longbibliography,twocolumn]{revtex4-1}

\usepackage[USenglish]{babel}

\usepackage{amsmath,amssymb,amsfonts}

\usepackage{bm}

\usepackage{epsfig}
\usepackage{subfigure}

\usepackage[colorlinks=true,linkcolor=blue,citecolor=red]{hyperref}

\newcommand{\equ}[1]
{Eq.~(\ref{#1})}

\newcommand{\figu}[1]
{Fig.~\ref{#1}}

\newcommand{\secu}[1]
{Sec.~\ref{#1}}

\def\e{\varepsilon}  
\def\z{\zeta}

\def\HH{{\cal H}}

\def\OO{{\cal O}}
\def\DD{{\cal D}}

\def\RRR{\mathbb{R}}

\def\=={\equiv}

\def\cG0{{\cal G}_0} 
\def\cG{{\cal G}}

\def\bra{\langle} 
\def\ket{\rangle}

 \def\Im{\mbox{Im}}

\def\=={\equiv}

\def\Im{{\rm Im}}

\newcommand{\eqn}[1]{(\ref{#1})}
\newcommand{\dagga}{{\phantom{\dagger}}}
\newcommand{\be}{\begin{equation}}
\newcommand{\ee}{\end{equation}}
\newcommand{\dis}{\displaystyle}
\newcommand{\fract}[2]{\frac{\dis #1}{\dis #2}}

\newcommand{\bk}{\mathbf{k}}
\newcommand{\bX}{\mathbf{X}}

\begin{document}

\date{\today}

\author{F.~Grandi}
\affiliation{Scuola Internazionale Superiore di Studi Avanzati
  (SISSA), Via Bonomea 265, I-34136 Trieste, Italy}
\affiliation{Department of Physics, University of Erlangen-N\"urnberg,
  91058 Erlangen, Germany}

\author{A.~Amaricci}
\affiliation{Scuola Internazionale Superiore di Studi Avanzati
  (SISSA), Via Bonomea 265, I-34136 Trieste, Italy}
\affiliation{CNR-IOM DEMOCRITOS, Istituto Officina dei Materiali,
  Consiglio Nazionale delle Ricerche, Via Bonomea 265, I-34136
  Trieste, Italy}

\author{M.~Fabrizio}
\affiliation{Scuola Internazionale Superiore di Studi Avanzati
  (SISSA), Via Bonomea 265, I-34136 Trieste, Italy}

\title{Unraveling the Mott-Peierls intrigue in Vanadium dioxide}

\begin{abstract}
Vanadium dioxide is one of the most studied strongly correlated 
materials. Nonetheless, the intertwining between electronic
correlation and lattice effects has precluded a comprehensive description 
of the rutile metal to monoclinic insulator transition, in turn
triggering a longstanding ``the chicken or the egg'' debate about which comes first,
the Mott localisation or the Peierls distortion.
Here, we suggest that this problem is in fact ill-posed: the electronic
correlations and the lattice vibrations conspire to stabilise the
monoclinic insulator, and so they must be both considered not
to miss relevant pieces of the VO$_2$ physics.
Specifically, we design a minimal model for VO$_2$  
that includes all the important physical ingredients: the electronic
correlations, the multi-orbital character, and the two components 
antiferrodistortive mode that condenses in the monoclinic
insulator. We solve this model by dynamical mean-field theory within
the adiabatic Born-Oppenheimer approximation. Consistently with the
first-order character of the metal-insulator transition, the
Born-Oppenheimer potential has a rich landscape, with minima
corresponding to the undistorted phase and to the four equivalent
distorted ones, and which translates into an equally rich
thermodynamics that we uncover by the Monte Carlo method.
Remarkably, we find that a distorted metal phase intrudes
between the low-temperature distorted insulator and high-temperature
undistorted metal, which sheds new light on the debated experimental
evidence of a monoclinic metallic phase.
\end{abstract}
\maketitle

\section{Introduction}\label{secI}
Vanadium dioxide (VO$_{2}$) is a transition metal compound with 
tremendous potential for technological applications, essentially in reason of its nearly room
temperature metal-to-insulator transition
\cite{doi:10.1146/annurev-matsci-062910-100347,Driscoll1518,smart_windows,doi:10.1063/1.3458706,doi:10.1002/smll.201802025,Nature_tera_meta,Nature_metadevice,PhysRevX.3.041004,Nature_firing,PhysRevApplied.11.054016}. 
Over the years, VO$_{2}$ has been subject to an intense
investigation, which dates back to the first decades of the last
century~\cite{doi:10.1002/zaac.19392420107,doi:10.1021/ja01194a051,andersson1,Nature_magn_susc,magneli1955,andersson2,doi:10.1002/zaac.19582970102,PhysRevLett.3.34,westman1961,Klemm1939},
but that is yet alive~\cite{Averitt-PRB2018,Wall572,Cheneaav6815} and,
to some extent,
debated~\cite{PhysRevLett.94.026404,PhysRevLett.107.016401,PhysRevLett.117.056402,PhysRevB.95.035113,PhysRevB.96.054111,Rozenberg-PRB2018,PhysRevB.87.195106}.
At the critical temperature $T_{c} \sim 340~\mathtt{K}$ and ambient
pressure, VO$_2$ undergoes a first-order transition from a metal
($T>T_c$) to an insulator
($T<T_c$)~\cite{nature12425,doi:10.1021/acs.nanolett.7b00233}, both
phases being paramagnetic
\cite{doi:10.1002/bbpc.19640681015,Villeneuve1985,PhysRev.185.1022}.
In concomitance with the metal-insulator transition, a structural distortion occurs from 
a high-temperature rutile (R) structure to a low temperature monoclinic 
(M1) one. 

\begin{figure}
 \centering \includegraphics[width=0.5\textwidth]{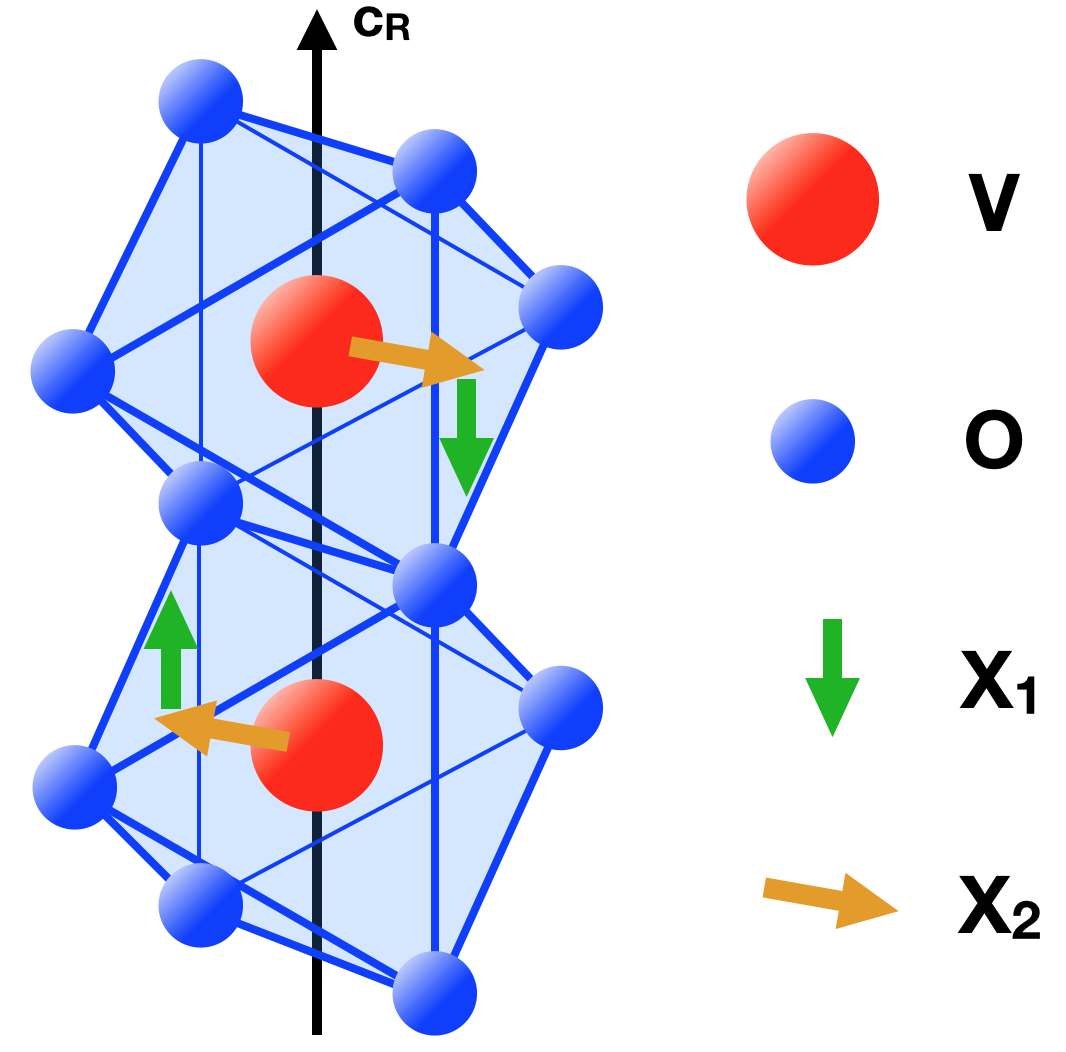}
  \caption{ (Color online) The rutile crystal structure, where the large (small) spheres represent Vanadium (oxygen) atoms. A cartoon of the $X_{1}$ and $X_{2}$ lattice distortions is also depicted, where the $X_{1}$ component acts as a dimerisation along the $c_{R}$ axis and the $X_{2}$ component acts as a tilting in the perpendicular plane. The monoclinic M1 phase is actually characterised by finite displacements both of $X_1$ and $X_2$.}
  \label{fig_rutile_struc}
\end{figure}

The crystal structure of rutile VO$_2$ is formed by  
equally spaced apart Vanadium atoms sitting 
at the centre of edge-sharing oxygen octahedra that form linear chains  
along the R $c$-axis, which we shall denote as c$_{R}$, see 
Fig.~\ref{fig_rutile_struc}. The tetragonal crystal field splits the $3d$-manifold 
into two higher e$_g$ and three lower t$_{2g}$ levels. In the oxidation state 
V$^{4+}$, the single valence electron of Vanadium can, therefore, occupy 
any of the three t$_{2g}$ orbitals, which are in turn distinguished
into a singlet $\text{a}_{1g}$ (or $d_{||}$) and a doublet
$\text{e}^\pi_g$ (or $d_{\pi^*}$), having, respectively, bonding and
non-bonding character along the c$_R$-axis. 
The M1 phase is instead characterised by an
anti-ferroelectric displacement of each Vanadium away from the centre 
of the octahedra, see Fig.~\ref{fig_rutile_struc}, so that the 
above-mentioned chains, from being straight in the R phase,
turn zigzag and dimerise
\cite{0957-4484-27-43-435704,Baum788}.  

A simple portrait of the transition in VO$_2$ was proposed in 1971
by Goodenough~\cite{GOODENOUGH1971490}. According to his proposal,  
the basal-plane component of the anti-ferroelectric 
distortion raises the energy of $\text{e}^\pi_g$ with respect to
the $\text{a}_{1g}$ \cite{condmat2040038}.
In addition, the c$_R$ component of the distortion, which drives the chain
dimerisation, opens a hybridisation gap between bonding and anti-bonding 
combinations of the $\text{a}_{1g}$. For large enough
crystal field splitting and hybridisation gap, the bonding combination
of the $\text{a}_{1g}$ fills completely, while the anti-bonding as
well as the $\text{e}^\pi_g$ get empty, hence the insulating
behaviour. 
The Goodenough's mechanism for the metal-insulator transition in
VO$_2$ thus relies on a single-particle description: the Peierls 
instability of the quasi-one-dimensional $\text{a}_{1g}$ band that
becomes half-filled after the growth of the crystal field drained 
the $\text{e}^\pi_g$ orbital.

However, Pouget et al.~\cite{PhysRevB.10.1801} and later Zylbersztejn and Mott~\cite{PhysRevB.11.4383} soon after argued
that the role of electronic correlation cannot be neglected
as in the Goodenough's scenario. Indeed, a tiny $\sim 0.2\%$
substitution of V with Cr changes the low-temperature insulator 
from the M1 crystal structure to a new monoclinic phase, named M2,
where dimerised and zigzag chains alternate \cite{pouget:jpa-00216522,Villeneuve1985}. 
The M2 phase can be also stabilised under hydrostatic pressure or uniaxial
stress~\cite{PhysRev.185.1034,PhysRevB.10.1801,Pouget-PRL-1975,nature12425,doi:10.1021/acs.nanolett.7b00233,PhysRevB.94.085105}.
In addition, a triclinic (T) phase with intermediate
structural properties~\cite{pouget:jpa-00216522} was shown to intrude
between M1 and M2.
The zigzag undimerised chains in M2 are still insulating and display 
magnetic properties akin those of a spin-$1/2$ 
antiferromagnetic Heisenberg chain~\cite{PhysRevB.10.1801,PhysRevB.5.2541,pouget:jpa-00216522}.
This likeness can be rationalised only invoking sizeable electronic
correlations.
Given the low concentration of substitutional Chromium or the small value of
uniaxial stress required to stabilise M2,  
it is reasonable to conclude that M1 must be as correlated as M2 
\cite{PhysRevLett.73.3042,Mott_Friedman,0953-8984-15-19-322,doi:10.1080/14786437008238476}. 

We believe that, even though electronic correlations are likely
necessary, they are nonetheless not sufficient to explain the phase
diagram of VO$_2$. It is known that a strong enough repulsion may
drive a Mott transition in a three-band Hubbard model at the density of
one electron per site~\cite{Werner-PRL2008}. Therefore, it is well
possible that the insulating phase of VO$_2$ is driven by correlations
alone, and that the structural distortion below $T_c$ is just the best
way the Mott insulator can freeze the residual spin and orbital
degrees of freedom to get rid of their entropy. However, should that
be the case, VO$_2$ would most likely remain insulating even above
$T_c$, which is not the case, all the more so because  
$k_{B} T_c$ is more than one order of 
magnitude smaller than the optical gap in the M1
phase~\cite{PhysRevB.77.115121}. For the same reason, we must exclude  
a transition merely driven by the larger electronic entropy of
the metal.  

We are thus inclined to believe that the structural distortion is also
necessary to stabilise the insulating phase in VO$_2$, but, once
again, not sufficient in view of the behaviour of the M2 phase, and of
the \textit{bad metal} character of the R
phase~\cite{Lee-Science2017,Qazilbash-Science2007,PhysRevB.74.205118}.
It is therefore quite likely that Goodenough's scenario is after all
correct, though it requires an active contribution from electronic
correlations. 

Indeed, different DFT-based calculations, which should properly
account for the effects of the lattice distortion on the electronic
structure, though within an independent-particle scheme, do not agree one with
another, and none explains at once all experiments.
For instance, straight LDA or GGA 
methods do not find any gap opening in M1 and M2 phases
\cite{doi:10.1002/1521-3889,PhysRevB.71.085109}. 
Such gap is instead recovered by GW~\cite{PhysRevB.60.15699,PhysRevLett.99.266402,Sakuma_2009} 
or LDA+U~\cite{Korotin2003,Kozhevnikov2007,PhysRevB.86.235103}, in all 
its variants. However, GW does not give easy access 
to the total energy, and therefore it does not explain why low temperatures 
should favour the M1 distorted phase against the rutile undistorted one.  
In turns, LDA+U or GGA+U calculations, known to overemphasise the exchange 
splitting, predict the existence of local moments even in the rutile 
phase~\cite{Korotin2003,Kozhevnikov2007,PhysRevB.86.235103}, 
not observed in experiments~\cite{Pauli-susc}.
Relatively recent calculations based on HSE hybrid functionals bring even worst 
results: both rutile and M1 phases are predicted to be magnetically ordered 
insulators, with the former lower in energy~\cite{PhysRevB.86.081101,PhysRevB.88.041107}, 
even though earlier calculations were claimed to be more in accordance 
with experiments \cite{PhysRevLett.107.016401}.
In turn, mBJ exchange potentials seem to predict the proper 
conducting behaviour of the R and M1 phases, as well as their lack of magnetism 
\cite{PhysRevB.86.075149}, which is erroneously predicted to occur also in the 
M2 phase~\cite{PhysRevB.86.235103}. This suggests that suppression of magnetic 
moments is somehow the rule of mBJ functionals applied to VO$_2$, which only by 
chance is the correct result for R and M1 phases.    
Finally, calculations based on PBE0 hybrid functionals properly account for the magnetic and
electronic properties of M1 and M2 phases, but predict ferromagnetism
in the rutile structure, at odds with
experiments~\cite{doi:10.1002/bbpc.19640681015},
as well as the existence of a never observed ferromagnetic and insulating monoclinic phase,  
dubbed M0 \cite{PhysRevB.95.125105}, also predicted by 
PBEsol functionals \cite{Moatti2019}.

One might expect that combining \textit{ab-initio} techniques with
many-body tools, e.g., DFT with dynamical mean-field theory (DMFT)
\cite{Kotliar2006a}, should work better and finally provide
uncontroversial results in accordance with experiments.
Unfortunately, different calculations by state-of-the-art DFT+DMFT methods do not even agree
about an unanimous view of the M1 monoclinic phase. Specifically, M1
has been regarded from time to time as a correlation-assisted Peierls
insulator~\cite{PhysRevLett.94.026404,PhysRevB.85.045109}, or, vice
versa, as a Peierls-assisted Mott
insulator~\cite{PhysRevLett.108.256402}, or, finally, as a genuine
Mott
insulator~\cite{PhysRevB.73.195120,PhysRevLett.117.056402,PhysRevB.96.195102}.

In view of the above controversial results, we think it is worth 
desisting from describing VO$_2$ straight from first principles, and
rather focusing on a minimal model, which can include all
the ingredients that are, by now, widely accepted to be essential.
As we mentioned, electron-electron correlations must play an important
role and thus need to be included and handled in a truly many-body
scheme.  
At the meantime, the coupling of the electrons to the lattice is equally 
important and must be included as well. We
earlier mentioned that the monoclinic distortion in the M1 phase
actually entails two different antiferrodistortive components:
the basal-plane displacement of V from the octahedron centre,
resulting into a zigzag shape of the formerly straight chains, and the
out-of-plane displacement that produces the chain dimerisation. The
two phenomena may actually occur separately, as indeed proposed by
Goodenough~\cite{GOODENOUGH1971490}, who argued that, generically, the
basal-plane distortion should appear at higher temperatures than
dimerisation. Indeed, time-resolved spectroscopy measurements during a
photoinduced monoclinic-to-rutile transition have shown that dimerisation
melts on earlier time-scales than the basal-plane displacement
\cite{Baum788,Kumar-PRB2017,Yuan-PRB2013}, which therefore must be
distinct and actually more robust than the former. We must mention,
however, that this conclusion does not agree with other
experiments~\cite{Morrison445,Otto450,PhysRevLett.109.166406,PhysRevLett.105.056404,Otto450}.
More convincing evidence is offered by the monoclinic metal that intrudes,
under equilibrium conditions, between rutile metal and monoclinic
insulator at ambient pressure
\cite{doi:10.1143/JPSJ.21.1461,Laverok-PRL2014,Yao-PRL2010,Lee1037},
and nor just above a critical pressure as originally believed
\cite{PhysRevLett.98.196406}. This phase might correspond to a crystal
structure where dimerisation is almost melted unlike the zigzag
distortion \cite{Yao-PRL2010,Moatti2019}, so that $\text{e}^\pi_g$ are
still above the $\text{a}_{1g}$, though the dimerisation is too
weak to stabilise at that temperature an hybridisation gap within the
$\text{a}_{1g}$ band~\cite{PhysRevB.95.035113}. Even the disappearance
prior to the metal-insulator transition \cite{Gray-PRL2016} of the
so-called singlet peak, which is  
associated to dimerisation and observed in optics, can be regarded as
a consequence of the melting of dimerisation preceding the complete
monoclinic-to-rutile transformation.  
All the above experimental facts point to the need to treat separately the basal-plane 
displacement and the out-of-plane one.
Finally, the importance of the basal plane antiferrodistortive
mode suggests the last ingredient to be considered: the
multi-orbital physics. This aspect was originally emphasised by
Goodenough~\cite{GOODENOUGH1971490} and successively confirmed by
many optical
measurements~\cite{PhysRevB.77.115121,Nature_orb,PhysRevB.95.075125}.

To summarise, we shall consider a microscopic model which includes 
the following relevant features:
\begin{enumerate}
\item{the electron-electron correlations and the coupling to the lattice distortion \cite{PhysRevB.5.2541,Nature_ph,Lee-Science2017,PAUL1970691,LEDERER19721969,PhysRevLett.114.116402,Hwang2017,PhysRevB.7.326,PhysRevB.10.490,PhysRevLett.27.727,refId0,0022-3719-1-1-322,PhysRevB.16.3338,0953-8984-12-41-310,PhysRevB.60.13249,PINTCHOVSKI1978941,Nat_comm_Wall,PhysRevLett.99.116401,WEGKAMP2015464,SCHILBE2004449,PhysRevLett.27.727,PhysRevB.1.2557,PhysRevLett.87.237401,PhysRevB.93.115126,2020arXiv200108580M};}
\item{the existence of two different antiferrodistortive components, each playing its own
distinctive role \cite{GOODENOUGH1971490,Baum788,Kumar-PRB2017};}
\item{the multi-orbital physics \cite{GOODENOUGH1971490,PhysRevB.77.115121,Nature_orb,PhysRevB.95.075125}.}
\end{enumerate}
with the minimal requirement of capturing, at
least at a qualitative level, the following aspects of the VO$_2$ physics:
\begin{itemize} 
\item[A.] the existence of an undistorted paramagnetic metal and a monoclinic distorted insulator \cite{Scirep_res,PhysRev.117.1442,PhysRev.185.1034,PhysRevLett.17.1286,ADLER19681};
\item[B.] the first-order character of the transition between them \cite{doi:10.1143/JPSJ.19.1748,LADD1969425,doi:10.1063/1.1656687,BONGERS1965275,PhysRevLett.3.34,doi:10.1021/nl1032205,Nature_coll_deloc,doi:10.1063/1.4861901,doi:10.1143/JPSJ.21.819B,doi:10.1143/JPSJ.19.517,CHANDRASHEKHAR1973369,1347-4065-6-9-1060};
\item[C.] the possible existence of an intermediate monoclinic metal \cite{doi:10.1143/JPSJ.21.1461,PhysRevB.77.235401,doi:10.1063/1.4764040,Yao-PRL2010,Laverok-PRL2014,doi:10.1021/nl9028973,PhysRevLett.97.266401,PhysRevB.85.155120,doi:10.1002/adma.201402404,Ilinskiy2012a,Ilinskiy2012b,doi:10.1063/1.3009569};
\end{itemize} 

Many models have been already put forth to describe VO$_2$. 
However, most of them focus either on the role of the
electron-electron correlations, or on that of the electron-lattice
coupling~\cite{PhysRevLett.25.376,0022-3719-5-12-012,PhysRevB.78.115103,PhysRevB.95.035113,Rozenberg-PRB2018,doi:10.1143/JPSJ.20.516,HEARN1972447,HYLAND1970318,PhysRev.155.851,MITRA1971221,CHATTEJEE197256,PhysRevB.7.3753,PhysRevB.7.3760,WOODLEY2008167,doi:10.1021/cm701861z,Nano_lett_GL},
and thus do not allow accessing in a single framework the whole VO$_2$ 
phase diagram, e.g., the points A., B. and C. above. 
Despite that, we must mention that the purely electronic Dimer Hubbard Model presented in \cite{PhysRevB.95.035113}, which by construction is not able to capture the monoclinic to rutile phase transition, is nevertheless able to describe some of the observed features of the monoclinic metal, like the MIR  peak in the optical conductivity observed in \cite{Qazilbash-Science2007}. 
There are actually some exceptions where electron-electron and
electron-lattice interactions have been considered on equal
footing~\cite{deGraaf,PhysRevB.22.5284,SHI19913527}.
In particular, the model studied in~\cite{PhysRevB.22.5284} includes explicitly all ingredients listed above. However, therein it is assumed a small bandwidth of the $\text{a}_{1g}$-derived band as compared to the $\text{e}^\pi_g$ one, which contradicts LDA calculations~\cite{doi:10.1002/1521-3889}. Moreover, \cite{PhysRevB.22.5284} includes the two distinct effects of the monoclinic distortion, but parametrized by a single displacement variable. In this way they preclude the possibility to describe the emergence of the monoclinic metal that seems to be observed experimentally. Furthermore, the mean-field treatment of the electron-electron interaction, despite its strength being comparable to the conduction bandwidth, yields not surprisingly to the formation of local moments in the rutile metal, not in accordance with magnetic measurements~\cite{Pauli-susc}. This negative result, highlighted by the same authors of Ref.~[\onlinecite{PhysRevB.22.5284}], solicits for a more rigorous treatment of the interaction.


This is actually the scope of the present work, which is organised as
follows. In \secu{secII} we introduce a simple model that includes the
three ingredients previously outlined, which we believe should capture
the main physics of Vanadium dioxide. In \secu{secIII} we 
discuss the dynamical mean-field theory (DMFT) approach to the model 
Hamiltonian, and presents in \secu{subsecIII.A} its ground state phase 
diagram. In \secu{secIV} we discuss the insulator-metal transition that occurs 
in our model upon raising the temperature. In \secu{subsecIV.A} we discuss the 
case in which such transition is driven solely by the electronic entropy, hence 
neglecting the lattice contribution to entropy, whereas 
in \secu{subsecIV.B} the opposite case. We will show that the latter situation 
is rather suggestive, since it foresees different transition temperatures 
of the two antiferrodistortive 
components, as predicted by Goodenough~\cite{GOODENOUGH1971490}. 
In turn, this result might explain the evidence supporting the existence 
of a monoclinic metal phase. 
Finally, \secu{secV} is devoted to concluding remarks. 

\section{The model}\label{secII}
As we mentioned, the orbitals that are relevant to describe the physics of VO$_2$ are 
the Vanadium $3d-\text{t}_{2g}$ ones, comprising the $\text{a}_{1g}$ singlet and 
$\text{e}^\pi_g$ doublet, which host a single conduction electron. 
We believe that in this circumstance the doublet nature of the
$\text{e}_{g}^{\pi}$ is not truly essential; what really matters is the 
distinction between $\text{a}_{1g}$ and $\text{e}^\pi_g$ based on their bonding 
character with the ligands and response to atomic displacement. Therefore, in order 
to simplify our modelling without spoiling the important physics, we shall associate the 
$\text{e}_{g}^{\pi}$ doublet with just a single orbital~\cite{PhysRevB.87.205108,0022-3719-5-12-012}, 
which, together with the other orbital mimicking 
the $\text{a}_{1g}$ singlet, give rise to two bands, 
$\text{band~1}\leftrightarrow \text{a}_{1g}$ and 
$\text{band~2}\leftrightarrow \text{e}^\pi_{g}$, which accommodate one electron per site, 
i.e., they are quarter filled. 

The other ingredient that is necessary to properly describe VO$_2$ is the 
electron-electron Coulomb interaction. However, since the main role that 
Coulomb repulsion is believed to play is to suppress charge fluctuations on 
$V^{4+}$, we shall ignore the long range tail and replace Coulomb repulsion 
with a short-range interaction.  

Finally, we need to include the coupling to the lattice. For simplicity, we 
shall focus our attention only on the rutile and monoclinic 
M1 phases, as such ignoring the M2 phase, which is actually regarded by some as just a 
metastable modification of the M1 structure 
\cite{PhysRevB.10.1801,Pouget-PRL-1975,WOODLEY2008167}. 
Under this assumption, we can model the lattice antiferrodistortion 
through a two-component 
zone boundary mode at momentum $\mathbf{Q}$, with displacement $\mathbf{X}=(X_1,X_2)$ 
and classical potential energy $\Phi\big(X_1,X_2\big)$. The $X_{1}$ and $X_2$ 
components model, respectively, the dimerising out-of-plane displacement 
and the band-splitting basal-plane one, see \figu{fig_rutile_struc}
\cite{0953-8984-22-41-415601,PhysRevLett.25.376}. 

The model Hamiltonian is thus written as the sum of three terms:  
\begin{equation} \label{hamiltonian}
\mathcal{H} = \mathcal{H}_\text{el} + \Phi\big(X_1,X_2\big) + \mathcal{H}_{\text{el}-\mathbf{X}} \;. 
\end{equation}
$\mathcal{H}_{el}$ is the purely electronic component reading:  
\begin{equation}
\begin{split}\label{hamiltonian_el}
\mathcal{H}_\text{el} &= \sum_{a=1}^2\,\sum_{\bk}\,\big(\epsilon_{a\bk}-\mu\big)\,n_{a\bk}
+ \frac{U}{2} \sum_{i} n_{i} \left( n_{i} -1 \right)\,,
\end{split}
\end{equation}
where $n_{a,\bk}$ is the occupation number at momentum $\bk$ of the
band $a =1,2$, $n_{i}$ the electron number operator at site $i$, $\mu$
the chemical potential used to enforce the quarter filling condition and, finally, $U$
is the on-site Hubbard repulsion.

With the aim to reduce the number of independent Hamiltonian 
parameters, we assume that the density-of-states (DOS) 
$\mathcal{D}_{1}(\epsilon)$ and $\mathcal{D}_{2}(\epsilon)$, 
of the band 1 and 2, respectively, have same bandwidth and 
centre of gravity, which we shall take as the zero of energy. In 
addition, we consider both DOS symmetric with respect to their 
centre, and such that $\epsilon_{1\bk}=-\epsilon_{1\bk+\mathbf{Q}}$, 
where $\mathbf{Q}$ is the wave-vector of the antiferrodistortive 
mode $\mathbf{X}$. This assumption actually overestimates the 
dimerisation strength, since it entails that any $X_1\not=0$ is able to 
open a hybridisation gap in the middle of band 1, which, we remark, 
does not coincide with the chemical potential unless band 2 is 
pushed above it. This implies that a finite hybridisation gap within 
band 1 does not stabilise an insulator so long as band 2 still crosses 
the Fermi energy. Therefore our simplified modelling  does not spoil 
the important feature that a distorted insulating phase 
may occur only above a critical threshold of the Hamiltonian 
parameters, although it affects the value of that threshold, 
whose precise determination is however behind the scope of the present 
model-study.

\begin{figure}
 \centering \includegraphics[width=0.5\textwidth]{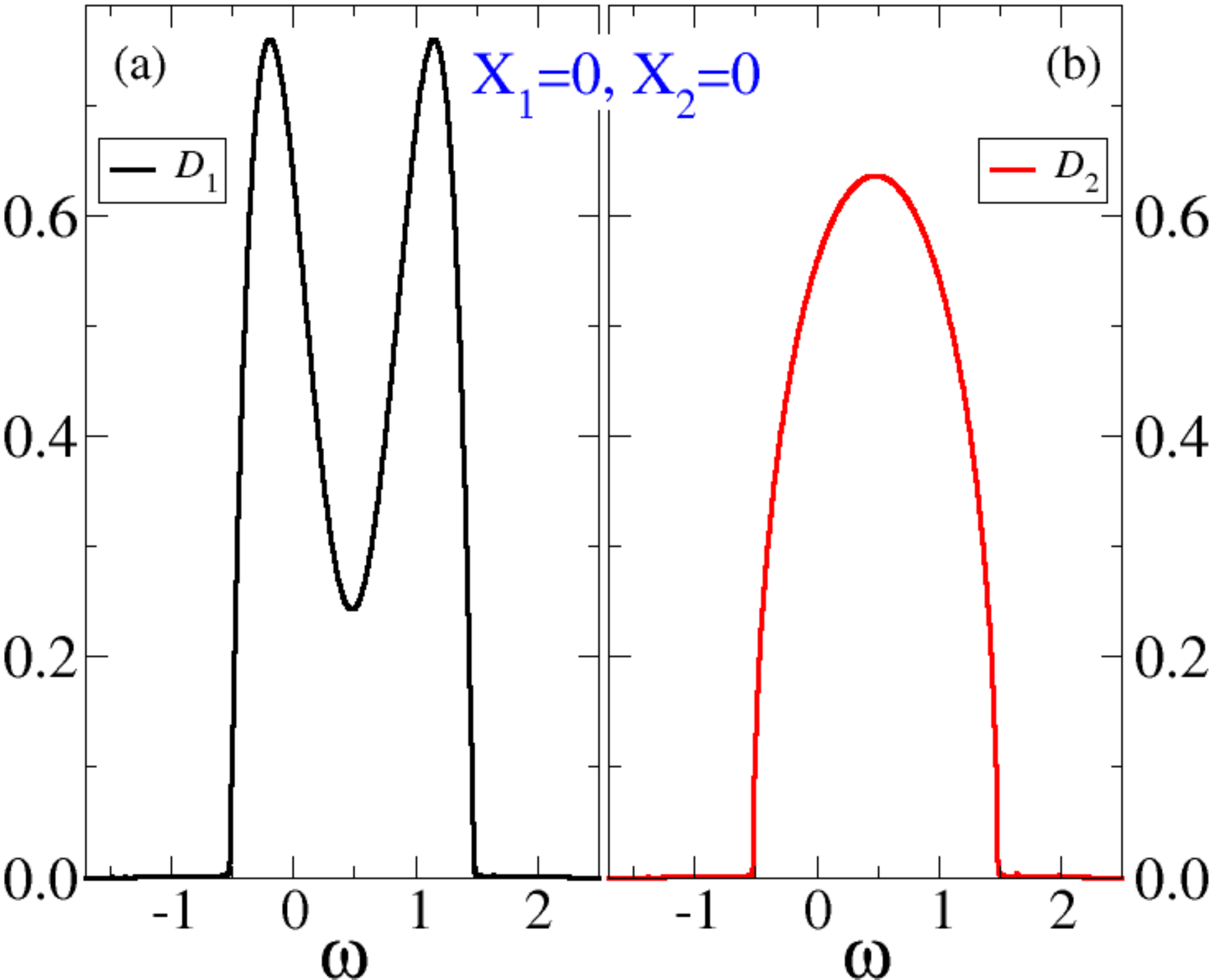}
  \caption{ (Color online) The density-of-states $\mathcal{D}_{a} \left( \omega \right)$, 
  a$=1,2$ for the two orbitals for $U=0$, $X_{1} = 0$ and $X_{2} = 0$.}
  \label{fig_dos_u0_mod}
\end{figure}

In order to emphasise the bonding character of the $\text{a}_{1g}$, band 1, along 
the $c_R$ axis, as opposed to the more isotropic $\text{e}^\pi_g$, band 2, we 
choose the following forms of the two corresponding DOS's: 
\begin{equation} \label{dos_eq}
\begin{split}
& \mathcal{D}_{1} \left( \epsilon \right) =  \frac{1}{\mathcal{N}} 
\Big[ a \epsilon^{2} -b \epsilon^{4} + D^{2} \big( b D^{2} -a \big) \Big] \,,\\
& \mathcal{D}_{2} \left( \epsilon \right) = \frac{2}{\pi D}\; \sqrt{1-\bigg( \frac{\epsilon}{D} \bigg)^{2}\;}
\;,
\end{split}
\end{equation}
with $\epsilon \in \left[ -D, D \right]$ and $\mathcal{N}$ a normalisation factor.  We take 
$b>a/D^{2}>0$ so that $\mathcal{D}_{1} \left( \epsilon \right)$ has a double-peak 
structure evocative of a one-dimensional DOS \cite{PhysRevB.22.5284,PhysRevLett.95.196404,PhysRevB.85.045109}. 
Hereafter, we take the half bandwidth $D=1$ as our energy unit, and fix 
$a D^{3} = 1.9$ and $b D^{5} = 2.1$. The resulting DOS's are shown in 
\figu{fig_dos_u0_mod} (a) and (b). There we note the two-peak structure of the band 1 
DOS, mimicking the Van Hove singularities of a quasi one-dimensional band structure, in 
contrast to the structureless band 2 DOS. 

We highlight that the electron-electron interaction in 
Eq.~\eqn{hamiltonian_el} only includes the monopole Slater integral 
$U>0$, and not higher order multipoles responsible of Hund's rules. 
This approximation, that makes the analysis more transparent, requires some justification. 
The Coulomb interaction of a single Vanadium projected onto the 
t$_{2g}$ manifold, which effectively behaves as an $l=1$ atomic shell, 
can be written in terms of two Slater integrals as: 
\begin{equation} 
\begin{split}
\mathcal{H}_{\text{t}_{2g}} &= \fract{U_{\text{t}_{2g}}-3J_{\text{t}_{2g}}}{2}\;
n^2  \\
&\qquad -\fract{J_{\text{t}_{2g}}}{2}\,\Big(4S(S+1) + L(L+1)\Big)\,,
\end{split}
\end{equation}
where $n$, $S$ and $L$ are the total occupation, spin, and angular momentum, respectively. 
Common values of the parameters are $U_{\text{t}_{2g}}\simeq 4~\text{eV}$ and 
$J_{\text{t}_{2g}}\simeq 0.68~\text{eV}\simeq U_{\text{t}_{2g}}/6$ \cite{PhysRevLett.94.026404}. 
Denoting as $E_0(n)$ the lowest energy with $n$ electrons in the t$_{2g}$ shell, 
the effective Hubbard $U$ for V$^{4+}$ can be defined through: 
\be
\begin{split}
U &= E_0(0)+E_0(2)-2E_0(1) = U_{\text{t}_{2g}}-3J_{\text{t}_{2g}} \simeq 1.96~\text{eV}
\,,
\end{split}
\ee
to be compared with the VO$_2$ bandwidth of about $2.6~\text{eV}$ 
\cite{doi:10.1002/1521-3889}. In units of the half-bandwidth, $U\simeq 1.5$, 
the value we shall use hereafter \cite{PhysRevB.86.165124,PhysRevLett.69.1796}. 
We observe that the Coulomb exchange $J_{\text{t}_{2g}}$ has no effect on the configurations 
with $n=0,1$, while it splits those with $n=2$ in three multiplets, with $(S,L)=(0,0),(1,1),(0,2)$, 
which are spread out over an energy $\simeq J_{\text{t}_{2g}}$, about a quarter of the full
bandwidth. Such small value is not expected to qualitatively alter the physical
behaviour, see, e.g., \cite{DeMedici2011}, which justifies our neglect of the exchange splitting 
in the model Hamiltonian \eqn{hamiltonian_el}.

We model the potential energy $\Phi\big(X_1,X_2\big)$ using a Landau functional for 
improper ferroelectrics \cite{0038-5670-17-2-R05,Kumar-PRB2017,doi:10.1002/1521-3889} 
expanded up to the sixth order in the lattice displacements:
\begin{equation} \label{hamiltonian_ph}
\begin{split}
 \Phi\big(X_1,X_2\big) & = N \bigg[\,\fract{\alpha}{2} \, \Big( X_{1}^{2} + X_{2}^{2} \Big) + \fract{\beta_1}{4} \,\Big( 2 X_{1} X_{2} \Big)^{2} \\
& \qquad\quad + \frac{\beta_2}{4} \Big( X_{1}^{2} - X_{2}^{2} \Big)^{2} + 
\frac{\gamma}{6} \Big( X_{1}^{2} + X_{2}^{2} \Big)^{3} \;\bigg] \,,
\end{split}
\end{equation}
where $N$ is the number of sites and the couplings $\alpha$ to
$\gamma$ are all positive. The terms proportional to $\alpha$, i.e. the
harmonic part of the potential, and that proportional to $\gamma$ have
full rotational symmetry in the $X_{1}$--$X_{2}$ plane.
On the contrary, $\beta_{1}$ favours a lattice distortion 
only along one of the two components, whereas $\beta_{2}$ a 
distortion with $|X_{1}| = |X_{2}|$.
In the specific case of VO$_2$, $\beta_2>\beta_1$, and thus it is preferable 
to equally displace both modes \cite{Kumar-PRB2017} rather than just one of them. 

Finally, we write the electron-lattice coupling as:
\begin{equation}\label{hamiltonian_el_ph}
\begin{split}
\mathcal{H}_{\text{el}-\mathbf{X}}  & = \mathcal{H}_{\text{el}-\mathbf{X}} \left[ X_{1},X_{2} \right] \\ 
& = -\fract{g}{2}\, X_{1} \sum_{\bk \sigma} \Big( c^\dagger_{1\bk\sigma}c^\dagga_{1\bk+\mathbf{Q}\sigma} + \text{H.c.} \Big) \\
& \ \ \ - \fract{\delta}{2}\,X_{2}^{2}\, \sum_\bk\,\Big(n_{1\bk}-n_{2\bk}\Big) 
\,,
\end{split}
\end{equation}
where $c_{1\bk\sigma}$ creates an electron at momentum $\bk$ 
in orbital $1$ with spin $\sigma$, and we recall that, by construction,  
$\epsilon_{1\bk}=-\epsilon_{1\bk+\mathbf{Q}}$. The dimerisation 
induced by the out-of-plane displacement $X_1$ is controlled by the 
coupling constant $g$, while $\delta$ parametrises the strength of the 
crystal field splitting generated by the basal-plane displacement 
$X_2$. By symmetry, the coupling between the field $X_{1}$ and the electron dimerization is at leading order linear \cite{PhysRevLett.25.376,SHI19913527}.
The quadratic coupling in $X_2$ is intentional and has a 
physical explanation. Indeed, $X_2$ corresponds to the Vanadium 
displacement parallel to the diagonal of the rutile basal plane away 
from the centre of the Oxygen octahedron. As a result, the 
hybridisation between the $\text{e}^\pi_g$ and the Oxygen ligands 
closer to the new Vanadium position increases, whereas the 
hybridisation with the further Oxygens diminishes \cite{PhysRevB.22.5284}. At linear order in 
the V-displacement $X_2$, the two opposite variations of the 
hybridisation cancel each other, but, at second order, they add up to 
a net rise in energy of the $\text{e}^\pi_g$ level, hence the expression 
in \equ{hamiltonian_el_ph}. The Hamiltonian \equ{hamiltonian} is 
invariant under the transformations $X_{1/2} \rightarrow - X_{1/2}$, 
reflecting a Z$_2 \times$Z$_2$ (also known as $K_{4}$ or 
``Vierergruppe'') symmetry.

Despite the great simplification effort, the model Hamiltonian 
\equ{hamiltonian} has still several parameters to be fixed. 
We emphasise that our main goal is to reproduce qualitatively the 
physics of VO$_2$, without any ambition of getting also a quantitative 
agreement. Nonetheless, just to be sure not to explore a Hamiltonian 
parameter space completely detached from the real VO$_2$ compound, 
we choose parameters in line with the existing literature. 
We already mentioned our choice of $U=1.5$, in units of the 
half-bandwidth, which is in line with the value used in realistic 
calculations~\cite{PhysRevLett.94.026404,doi:10.1080/14786436908228049,PhysRevB.81.115117,PhysRevB.22.5284,Mott_Friedman,SOMMERS1978133}.
The other parameters involve the phonon variables. 
We shall choose  $g = 0.4$, $\delta = 0.2$, $\alpha = 0.155$, 
$\beta_{1} = 1.75 \cdot 10^{-3}$, $\beta_{2} = 2 \beta_{1}$ and 
$\gamma = 6.722 \cdot 10^{-4}$. Those values permit to reproduce the inter-band character of the band gap experimentally observed for the monoclinic insulator \cite{Nature_orb} and to obtain a size of it close enough to the experimental findings, see \secu{sec_spectral_funct} for further details on the spectral properties of the system. Moreover, we checked \textit{a posteriori} that we can reasonably reproduce the size of the electron-phonon interaction \cite{PhysRevB.69.165104,0022-3719-5-12-012,PhysRevB.87.235118} and the lattice energy change across the rutile-to-monoclinic transition \cite{PhysRevB.87.235118} as they were obtained in previous experiments or theoretical analysis. As a concluding remark, we point out that the direct experimental fits of the coupling constants is satisfactorily in agreement with previous estimations of the same \cite{Kumar-PRB2017,PhysRevB.96.075128}, further corroborating our choice of the parameter set.


\section{DMFT solution} \label{secIII}
\begin{figure}
 \centering \includegraphics[width=0.5\textwidth]{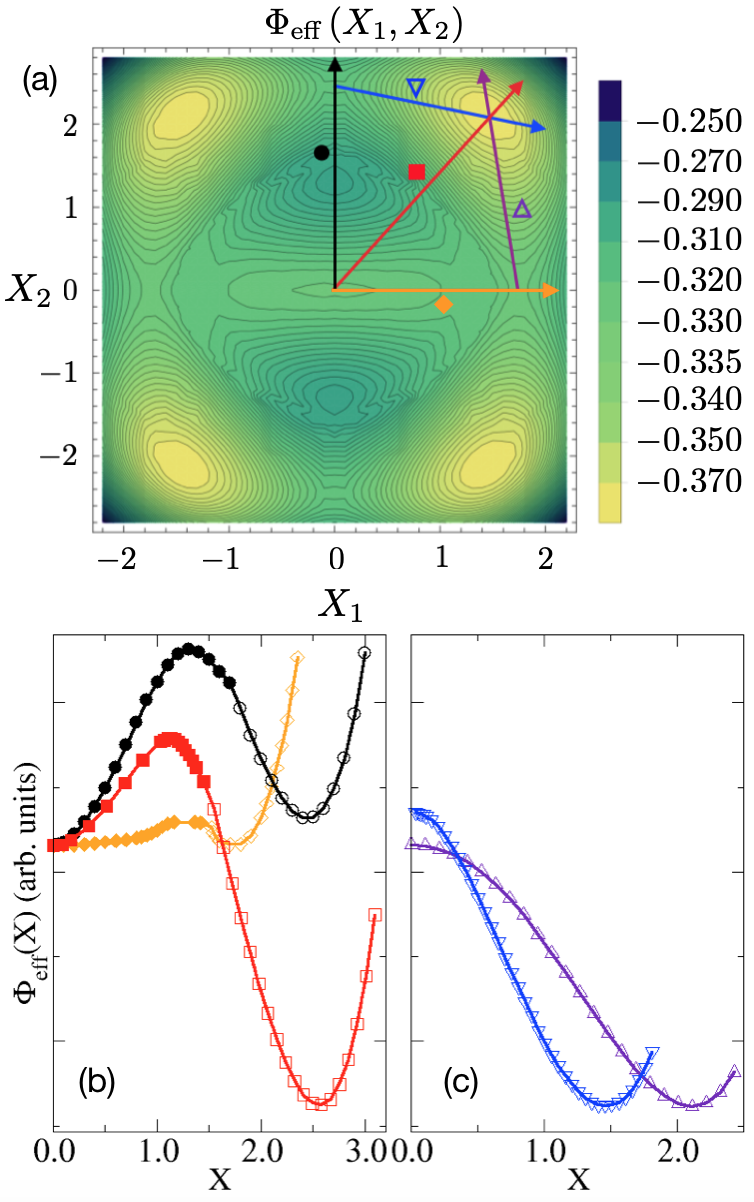}
  \caption{ (Color online) (a) The zero-temperature colour map of the
    internal energy of the system as function of the amplitude of the
    crystal distortions $X_{1}$ and $X_{2}$ for $U = 1.5$. The system
    displays five minima, one at $X_{1}=X_{2}=0$ corresponding to a
    metallic undistorted phase, the others at $X_{1} \simeq \pm 1.5$
    and $X_{2} \simeq \pm 2.1$ corresponding to four equivalent
    insulating and distorted states. (b),(c) Evolution of the zero
    temperature internal energy along the paths shown in panel (a),
    where the symbols close to them correspond to the ones used in
    panels (b) or (c); the coordinate $X = \sqrt{X_{1}^{2} +
      X_{2}^{2}}$ is computed along the lines as depicted in panel
    (a). Filled symbols correspond to a two band metallic solution,
    instead empty symbols correspond to an insulating solution
    everywhere except for the black curve with the circles, where they
    correspond to a single band metallic phase. Particularly, in panel
    (b): the circles (diamonds) correspond to the evolution of the
    internal energy along the line that involves just the distortion
    $X_{2}$ ($X_{1}$) and the squares correspond to the line that
    connects the undistorted metal and the distorted insulator found
    in the $ X_{1} $, $X_{2} > 0$ sector. In panel (c) the
    up-triangles (down-triangles) correspond to the line that connects
    the minimum observed in panel (b) along the line with the circles
    (diamonds) that involves just the distortion $X_{1}$ ($X_{2}$) to
    the global insulating minimum at $\left( X_{1}, X_{2} \right) =
    \left( 1.5, 2.1 \right)$.}
  \label{fig_en_vs_x1_x2}
\end{figure}

We solve the model Hamiltonian \equ{hamiltonian} by means of
DMFT~\cite{Georges1996RMP} within the adiabatic Born-Oppenheimer
approximation.
This approach will allow us to treat correlation effects
non-perturbatively beyond an independent-particle description.
Exploiting the Born-Oppenheimer approximation, we solve the electronic
problem at a fixed displacement $\mathbf{X}=(X_{1},X_{2})$. To any
choice of the displacement $\mathbf{X}$ it corresponds to a different
electronic problem through  the electron-phonon coupling discussed
above.
Within DMFT such resulting interacting lattice electrons problem is mapped onto a 
quantum impurity model constrained by a self-consistency condition, which
aim to determine the  bath so to describe the local physics
of the lattice model.
The effective bath is described by a frequency dependent Weiss field
$\mathcal{G}_{0,a}( i\omega_{n})$, with $a=1,2$ the
orbital index. The self-consistency condition relates the Weiss fields
to
the local self-energy function  $\Sigma_{a}(i\omega_{n})$, obtained from the solution of the effective quantum
impurity model, and the local interacting Green's function
$$
G_{\text{loc},a}(i\omega_{n} )=\int_\RRR d\e \DD_a(\e) \frac{1}{\z_{a}-\e} \;,
$$
where $\z_{a} = i\omega_n + \mu - \Sigma_a(i\omega_n)$. 
The self-consistency conditions read:
$$
\mathcal{G}^{-1}_{0,a} (i\omega_{n}) = G^{-1}_{\text{loc},a}(i\omega_{n}) + \Sigma_{a}(i\omega_{n}) \;.
$$

Once a Weiss field $\mathcal{G}_{0,a}(i\omega_n)$  is given, the
solution to the DMFT equations  is obtained iteratively as
follows.
We solve the effective quantum impurity problem associated to the
given Weiss field using Exact Diagonalization technique~\cite{Caffarel1994a,Weber2012PRB}. To this end,
we discretize the effective bath into a number $N_b$ of electronic
levels~\cite{Caffarel1994a,Capone2007PRB,Weber2012PRB}. The resulting
Hamiltonian is diagonalized using Lanczos method and the ground-state
(at zero temperature) or low lying states in the spectrum (at finite
temperature) are determined~\cite{Capone2007PRB}. The impurity Green's functions
$G_{a}(i\omega_{n})$ are then 
obtained using the dynamical Lanczos
technique~\cite{Georges1996RMP,Capone2007PRB}. The  
self-energy is obtained by solving the Dyson equation for the impurity
problem $\Sigma_a(i\omega_n) = \mathcal{G}^{-1}_{0,a} (i\omega_{n}) - G^{-1}_{a}(i\omega_{n})$.
The self-energy is used to evaluate the local interacting Green's
function and, finally, to update the Weiss fields by means of the
self-consistency condition.
The procedure is iterated until the overall error on the
determination of the Weiss field falls below a threshold, which in our
calculations was set to  $10^{-6}$.  
In this work, we use $N_b=8$ as the total number of bath sites,
corresponding to a finite system of $N=10$ electronic levels or
$N_s=20$ spins. Yet, we tested our results with respect to larger
values of $N_b$ without finding significant differences.

Using the DMFT method we computed the electronic properties for several values of the
displacement $\mathbf{X}=(X_{1},X_{2})$. The part of
\equ{hamiltonian_el_ph} related to the tilting $X_{2}$ enters in the
single-particle term of the impurity hamiltonian as the usual crystal
field splitting, while the dimerization $X_{1}$ acts directly on the
density of states of band $1$, see the expression of $ \mathcal{D}_{1}
\left( \epsilon \right)$ appearing in \equ{dos_eq}, and it opens a gap
of size $2 g X_{1}$ in correspondence to its center of gravity.

We calculate the total electronic energy, or the free-energy at finite 
temperature, which renormalizes the Born-Oppenheimer adiabatic 
potential of the displacement $\Phi ( X_{1},X_{2} )\to \Phi_\text{eff} ( X_{1},X_{2} )$ through:
\be
\Phi_\text{eff} ( X_{1},X_{2} ) = \Phi ( X_{1},X_{2} ) + \langle \,H_{\text{el}}\, \rangle + \langle \,H_{\text{el}-\mathbf{X}} \,\rangle\,.
\label{Phi_eff}
\ee
We shall restrict our analysis to the paramagnetic sector forcing spin 
$SU(2)$ symmetry. However, we did check that magnetic solutions are higher 
in energy. We first present results at zero-temperature $T=0$, and 
then move to those at  $T>0$.

\subsection{Ground state phase diagram}\label{subsecIII.A}

In \figu{fig_en_vs_x1_x2}a we show the adiabatic potential 
$\Phi_\text{eff} ( X_{1},X_{2} )$ in \eqn{Phi_eff} calculated by DMFT 
at $U = 1.5$. The energy landscape shows five minima. A local 
minima is located at the origin $X_{1} = X_{2} = 0$, and corresponds 
to an undistorted metal that we identify with the R phase of Vanadium 
dioxide. Four degenerate global minima are instead located at 
$X_{1}  \simeq \pm 1.5$ and $X_{2} \simeq \pm 2.1$, which are related 
to each other by the Z$_2\times$Z$_2$ symmetry and represent the 
four equivalent lattice distortions. We find that these global minima 
describe an insulating phase, and thus realize a two-band version of 
the Goodenough scenario~\cite{GOODENOUGH1971490} for the M1 
phase, in qualitative agreement with \textit{ab-initio} calculations of 
VO$_2$ ~\cite{WOODLEY2008167,doi:10.1021/cm701861z}. A 
detailed discussion of the electronic properties of all minima is 
postponed to the next \secu{sec_spectral_funct}.

In figures \ref{fig_en_vs_x1_x2}b and \ref{fig_en_vs_x1_x2}c we 
instead show the evolution of the adiabatic potential 
$\Phi_\text{eff} ( X_{1},X_{2} )$ along some specific lines, as indicated 
in \figu{fig_en_vs_x1_x2}a. We note that along the horizontal and 
vertical cuts, marked by a diamond and a circle in 
\figu{fig_en_vs_x1_x2}a, respectively, the energy landscape shows a 
saddle point, i.e., a minimum along the cut direction, but maximum in 
the perpendicular one. Within our model description, the effect of a 
uniaxial tensile strain would be taken into account by adding to the 
Hamiltonian \equ{hamiltonian} terms like: $-F_{1} X_{1}^{2}$ or 
$-F_{2} X_{2}^{2}$ ($F_{1}, F_{2} > 0$), depending  on the direction 
of the applied stress~\cite{PhysRevApplied.11.014054,Nat_nan_strain,ferroel}.
In presence of such terms, the saddle points observed in
\figu{fig_en_vs_x1_x2}a along the lines $X_{1} = 0$ or $X_{2} = 0$
may turn into additional minima of the energy
landscape~\cite{Nano_lett_GL}, which can possibly describe the occurrence of the M2 phase
in the framework of the same model Hamiltonian.

\begin{figure}
 \centering \includegraphics[width=0.5\textwidth]{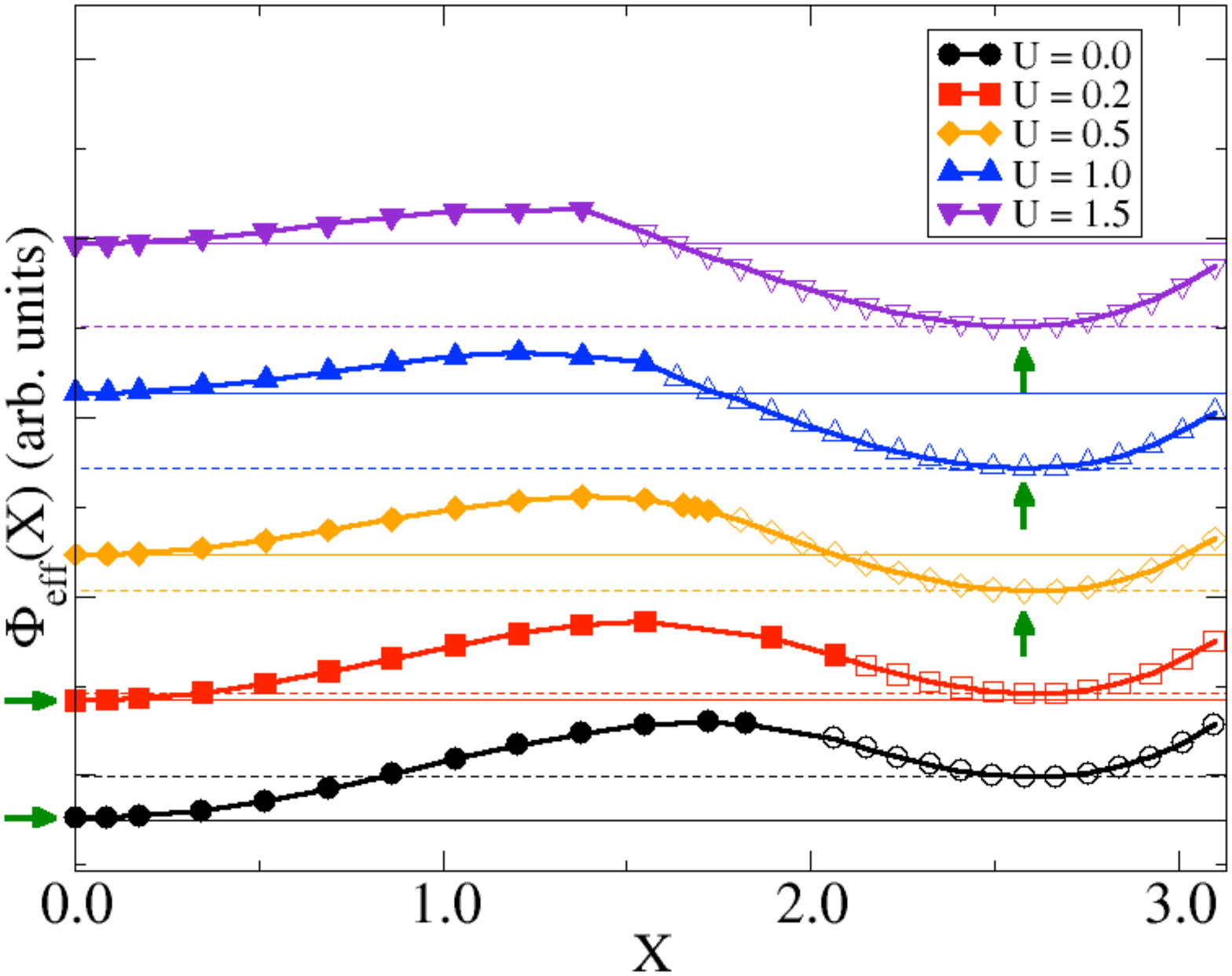}
  \caption{ (Color online) The zero-temperature internal energy of the system 
  (in arbitrary units) as function of the amplitude of the crystal distortion $X = \sqrt{X_{1}^{2} + X_{2}^{2}}$ (coordinate 
  taken along the line that connects the rutile solution and one of the monoclinic minima)
  for several values of the Hubbard interaction $U$. Filled (open) symbols 
  correspond to a metallic (insulating) solution. The continuous (dashed) horizontal lines 
  indicate the values of the metallic (insulating) minimum at each value of $U$. Arrows 
  indicate the position of the absolute minimum for each value of the interaction.}
  \label{fig_en_vs_x}
\end{figure}

In order to understand what is the role of the Hubbard interaction $U$
in stabilising the insulating solution, we studied the evolution of
$\Phi_\text{eff} ( X_{1},X_{2} )$ for several values of $U$, along the
line in the $X_{1}$--$X_{2}$ plane connecting the rutile
local minimum with one of the monoclinic global minima (the diagonal cut in Fig.~\ref{fig_en_vs_x1_x2}a marked by a diamond symbol). Our results are reported in \figu{fig_en_vs_x}. 
We note that already at $U=0$ the energy has two minima. One 
is at the origin and corresponds to the undistorted metal. The other is located at finite $\mathbf{X}$, and thus represents a distorted phase that must evidently be also insulating in order to be a local energy minimum. 
Therefore at small $U \lesssim 0.2$, the stable phase is
the undistorted metal at $X = 0$ in
\figu{fig_en_vs_x}, while the local minimum at $X\not=0$
(monoclinic insulator) is metastable.
However, for larger $U \gtrsim 0.2$, the situation is reversed: 
the distorted insulator becomes the global minimum, while the 
undistorted metal a local one, entailing the typical scenario of a first-order metal-insulator transition driven by interaction. 
The above results show that electron-electron interaction is crucial to 
stabilise the distorted insulator, though the active contribution of the 
lattice is equally essential. Indeed, the interaction strength, 
$U\simeq 1.5$ the half-bandwidth, is too small to drive on its own the metal-insulator transition \cite{DeMedici2011}. In other words, the picture that 
emerges from \figu{fig_en_vs_x}, with the interaction and the coupling to the lattice both necessary to stabilise the insulator,  
fully confirm our expectation in Sec.~\ref{secI}.  

\begin{figure}
 \centering \includegraphics[width=0.5\textwidth]{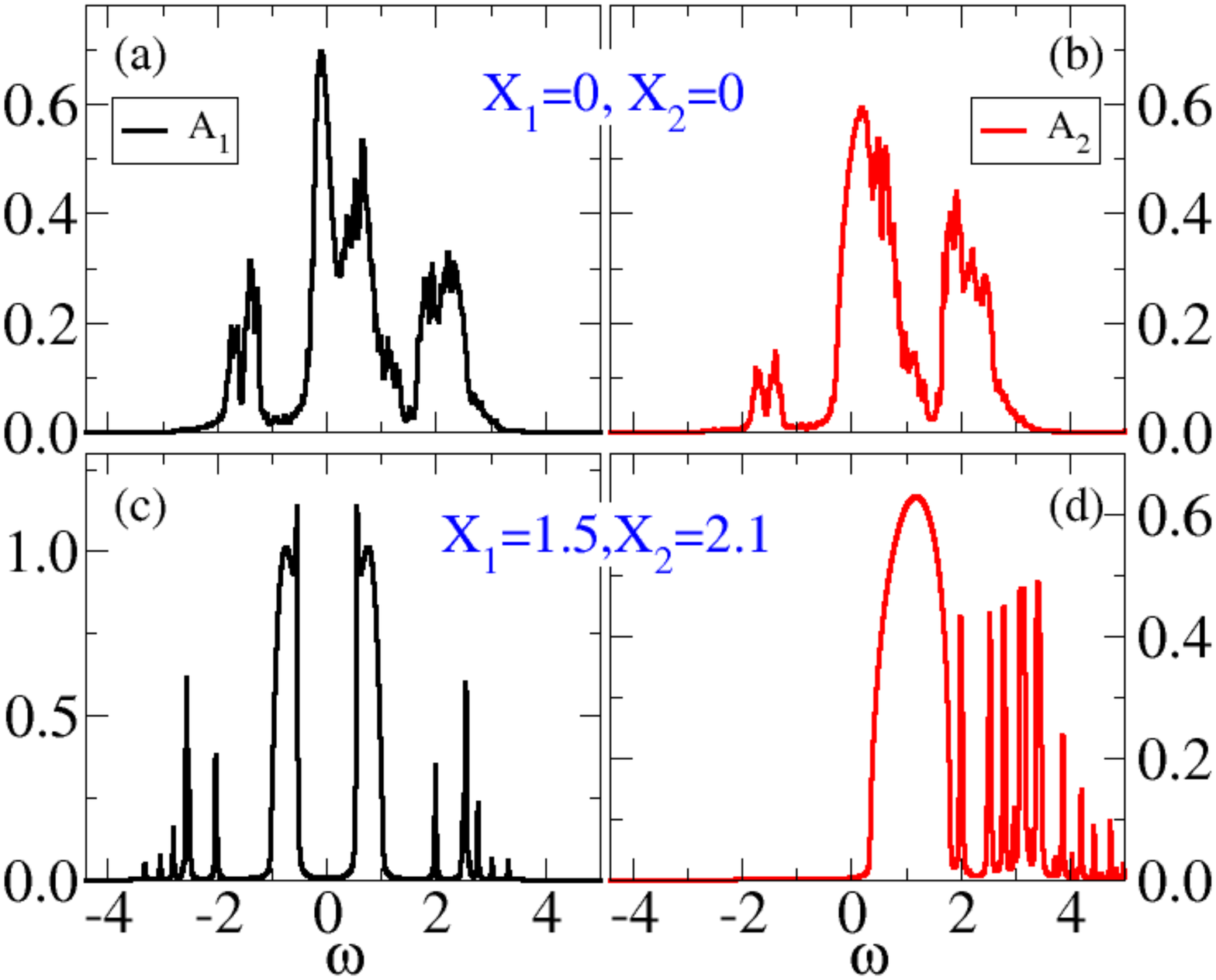}
  \caption{ (Color online) The spectral functions $A_{a} \left( \omega \right)$, 
  a$=1,2$ for the two minima shown in \figu{fig_en_vs_x} at $U=1.50$. The metallic 
  phase correspond to $X = 0$ [(a) and (b)], instead the insulator corresponds to 
  $X \sim 2.58$ [(c) and (d)].}
  \label{fig_dos_u1_50}
\end{figure}

\subsubsection{Spectral functions} \label{sec_spectral_funct}
Further insights into the properties of the metal-insulator transition
can be gained by looking at the spectral functions: 
\be
A_{a} \left( \omega \right) = - \frac{1}{\pi}\Im{G_{loc,aa}} \left( \omega \right)
\ee 
where $a = 1,2$ and $G_{loc,aa}$ is the local interacting Green's
function obtained within the DMFT solution of the model. 
In \figu{fig_dos_u1_50} we show $A_{a} \left( \omega \right)$  
at the different minima in \ref{fig_en_vs_x1_x2}a, with $\omega$
measured with respect to the chemical potential. 
We note that already in the absence of interaction, $U = 0$, the
different shapes of the DOS's, see \figu{fig_dos_u0_mod}, lead to a
larger occupation of band $1$ than band $2$. 
Such population unbalance is increased by $U>0$, which effectively enhances the crystal field, leading to an even larger
occupation of band $1$ at expenses of 
$2$~\cite{PhysRevB.87.205108,PhysRevB.91.115102,PhysRevB.98.045105}.
This is evident in the spectral function of the undistorted metal
at $U=1.5$, reported in \figu{fig_dos_u1_50}(a) and
\figu{fig_dos_u1_50}(b), where the occupied $\omega\leq 0$ part of 
$A_{1} \left( \omega \right)$ overwhelms that of 
$A_{2} \left( \omega \right)$ more than in the $U=0$ case of 
\figu{fig_dos_u0_mod}. 
We also note in the figures \ref{fig_dos_u1_50}(a) and
\ref{fig_dos_u1_50}(b) side peaks that correspond to the precursors of the Hubbard
bands. 

The scenario is radically different in the insulating solution, see
\figu{fig_dos_u1_50}(c) and \figu{fig_dos_u1_50}(d). Here we observe
the formation of a hybridisation gap opening at the chemical potential
inside the band 1. Two coherent-like features flank the gap.
The band 2 is instead pushed above the Fermi energy, and therefore is empty.
We still observe precursors of the Hubbard sidebands in $A_1(\omega)$, as well as signatures 
of the precursor of the upper Hubbard band in $A_2(\omega)$, though rather spiky
because of the bath discretisation. 

We note that in the insulating solution the lowest gap corresponds to 
transferring one electron from band 1 to 
band 2, i.e., from $\text{a}_{1g}$ to $\text{e}^\pi_g$ in the VO$_2$ language, 
and has a magnitude of about $E_\text{gap} \sim 0.8~\text{eV}$, for a realistic value of 
the half-bandwidth of 1.3~eV \cite{doi:10.1002/1521-3889}. 
This value of the gap is not too far from the experimental one, 
$E_\text{gap}^\text{ex} \sim 0.6-0.7~\text{eV}$ 
\cite{PhysRev.185.1022,PhysRevB.77.115121,doi:10.1021/nl1032205}. 
Therefore, our simplified modelling yields results that are not only qualitatively 
correct but, rather unexpectedly, also quantitatively not far off the actual ones. 
The band 1~$\to$~band 1 transition, i.e., $\text{a}_{1g}\to \text{a}_{1g}$, though 
being slightly higher in energy, has a much steeper absorption edge since it 
involves the two coherent peaks in \figu{fig_dos_u1_50}(c), already observed in previous works \cite{PhysRevLett.94.026404,PhysRevLett.117.056402,PhysRevB.81.115117,PhysRevB.95.035113}. This result is in 
loose agreement with XAS linear dichroism experiments 
\cite{PhysRevLett.97.116402,Gray-PRL2016} that are able to distinguish the 
two absorption processes.

In order to assess the degree of electronic correlations, we calculate
the quasiparticle residue of each band in the undistorted metal
phase, defined by: 
\be
Z_{a} = \Bigg(1- \fract{\partial \text{Re}
  \Sigma_{aa}(\omega)}{\partial\omega}\Bigg)^{-1} _{|\omega = 0}\;,
\ee
with $a = 1,2$.
We find that the two bands at $X_{1} = X_{2} = 0$ show almost the same 
value $Z_{1/2} \sim 0.67$, not inconsistent with more realistic calculations 
\cite{PhysRevB.81.115117,PhysRevLett.94.026404,Belozerov2011,PhysRevLett.108.256402}.
Such agreement, \textit{a priori} not guaranteed, gives further support to 
our simple modelling.

\section{Phase transition at finite temperature}\label{secIV}

Our main scope here is however to describe the first-order phase transition 
upon heating from the low-temperature M1 monoclinic insulator to the 
high-temperature rutile metal. In general, we can envisage a phase transition 
primarily driven either by the electron entropy or by the lattice one. 

Indeed, we note that the electron free energy of the metal solution, which is 
metastable at $T=0$, must drop faster upon raising temperature than the 
insulator free energy since the metal carries more electron entropy than the 
insulator. This effect alone, that is ignoring lattice entropy, would be able to 
drive a first-order transition when insulator and metal free energies cross.  
On the other hand, since the distorted ground state breaks the 
$Z_2\times Z_2$ symmetry of the adiabatic lattice potential $\Phi_\text{eff}(X_1,X_2)$ 
in \figu{fig_en_vs_x1_x2}, we might expect such symmetry to be recovered 
by raising temperature only because of lattice entropy effects, i.e., ignoring 
the electronic contribution to entropy.  

In reality, both effects should combine to drive the transition. However, dealing 
together with lattice and electron entropies within our computational scheme 
would imply to calculate the adiabatic potential $\Phi_\text{eff}(X_1,X_2)$ at 
any temperature, which is a rather heavy task. For this reason, in what follows 
we shall analyse separately electron and lattice entropy effects, and at the end 
argue what would happen should they act together.

\subsection{Electron-driven transition}\label{subsecIV.A}

Let us first neglect the lattice entropy and study the temperature evolution of 
the free energies of the two inequivalent minima that we found at zero 
temperature. For that, we need to evaluate the electronic entropy, which can 
be obtained through: 
\be \label{entropy}
  \begin{split}
    & S \left( X_{1}, X_{2}, T \right) = \int_{0}^{T} dT^{'} \ \frac{1}{T^{'}} \frac{\partial \Phi_{\text{eff}} \left( X_{1},X_{2}, T^{'} \right)}{\partial T^{'}} \\
    & = \int_{\Phi_{\text{eff}} \left( X_{1},X_{2}, 0 \right)}^{\Phi_{\text{eff}} \left( X_{1},X_{2}, T \right)} \frac{d \Phi_{\text{eff}}}{T^{'} \left( \Phi_{\text{eff}} \right)}\;.
  \end{split}
\ee
The last equality corresponds to a change of integration variable
from the temperature $T^{'}$ to the adiabatic potential $\Phi_{\text{eff}}$, which is also the internal energy.

\begin{figure}
 \centering \includegraphics[width=0.5\textwidth]{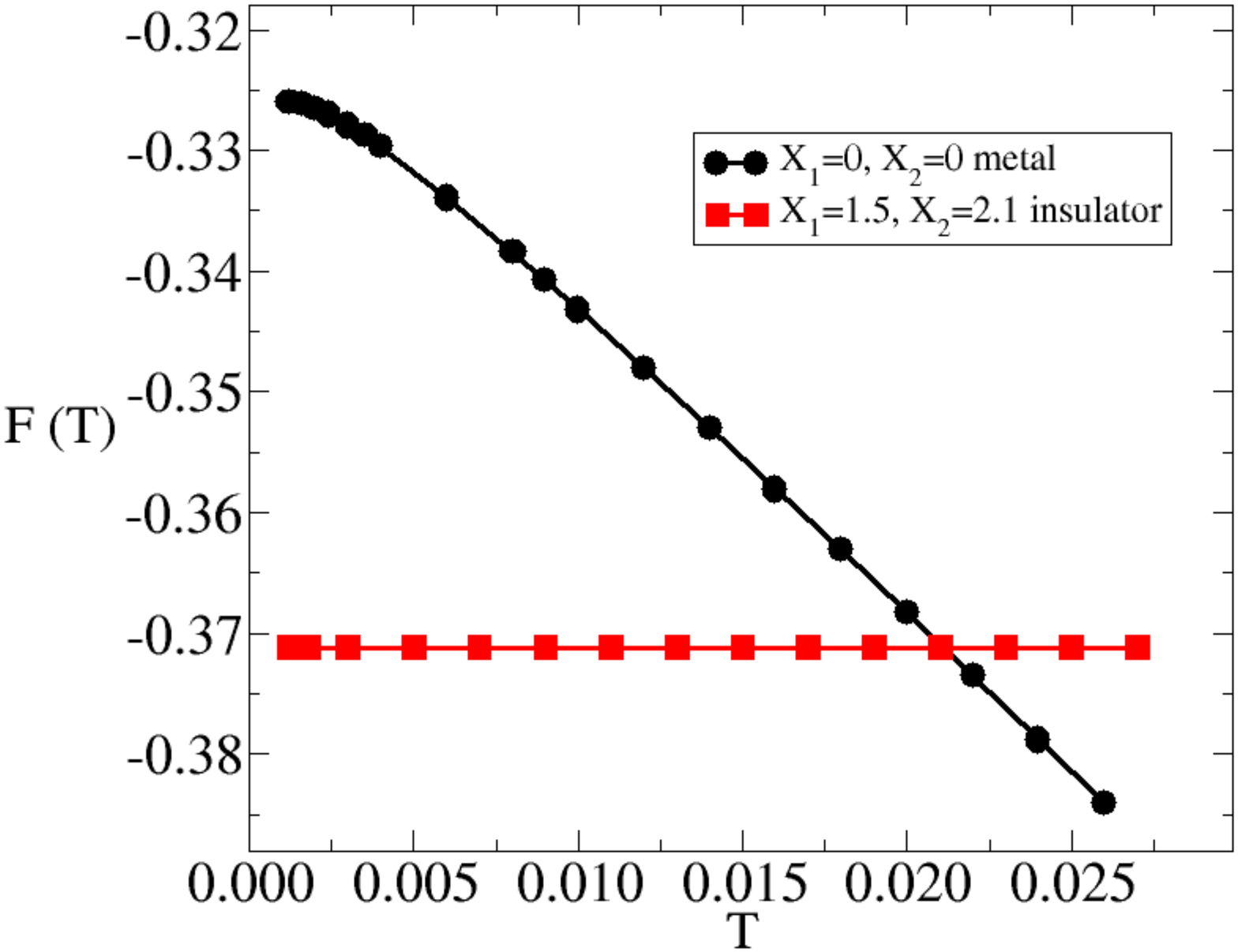}
  \caption{ (Color online) Temperature evolution of the free energy at the two inequivalent 
  minima $X_{1} = X_{2} = 0$ (dots) and $X_{1} = 1.5$, $X_{2} = 2.1$ (squares) observed at zero temperature 
  for $U=1.5$. The first-order transition occurs at $T_{\text{el}} \sim 0.021 \sim 320~\mathtt{K}$, of 
  the same order of magnitude as the experimental value $340~\mathtt{K}$.}
  \label{fig_free_en_vs_t}
\end{figure}

From the entropy $S$ we can estimate the free energy:
\be \label{free_en}
F\left( X_{1}, X_{2}, T \right) = \Phi_{\text{eff}} \left( X_{1}, X_{2}, T \right) - T  S \left( X_{1}, X_{2}, T \right)\,,
\ee
which, we emphasise once more, does not include the lattice contribution to 
entropy. We shall compare the free energy of the undistorted metal solution 
at $\mathbf{X}=0$, with that of the distorted insulator at $\mathbf{X}\not=0$. 
In principle, the equilibrium displacement in the insulator should change with 
temperature. In practice, since the entropy of the insulator is negligible for all 
temperatures under consideration, we shall fix $\mathbf{X}$ at the $T=0$ value. 
The temperature evolution of the metal and insulator free energies so obtained 
are shown in \figu{fig_free_en_vs_t}. As expected, the larger entropy of the 
metal pushes its free energy below the insulator one at relatively low 
temperature, $T_\text{el}\sim 0.021$, substantially smaller than the insulating 
gap, and thus justifying our assumption of frozen $\mathbf{X}$. $T_\text{el}$ 
identifies the insulator-metal transition, which is evidently first order since the 
two free energies cross with different slopes. Incidentally, $T_\text{el}\sim 0.021$ 
in half-bandwidth units, corresponds to $\sim 320~\text{K}$ for a realistic 
bandwidth of $2.6~\text{eV}$, which has the right order of magnitude 
when compared with the true critical temperature of $340~\text{K}$. However, we should mention that a different choice of the parameters appearing in \equ{hamiltonian}, keeping them still in a regime representative of the physics of Vanadium dioxide, would have produced a different value of $T_\text{el}$.



\subsection{Lattice driven transition} \label{subsecIV.B}

We now move to study the properties of the lattice-driven transition. For that, 
we first need to model the lattice dynamics. However, since the tetragonal R 
to monoclinic M1 transition is a complex structural transformation, with 
martensitic features, especially in films
\cite{martensitic-deAlmeida2000,martensitic-Lopez2002,martensitic-Pan2004,martensitic-Claassen2010,martensitic-Viswanath2013},
our modelling ought to be oversimplified, and aimed just to get qualitatively 
reasonable results, with no pretension of quantitative accuracy. 

As a first step, we must relax our previous assumption of a global 
antiferrodistortive mode, and instead introduce a displacement field, 
i.e., a site dependent displacement $\bX_i=\big(X_{1i},X_{2i}\big)$. We 
assume that $\bX_i$ feels the local adiabatic potential 
$\Phi_\text{eff}\big({\bf X}_{i}\big)$ of \figu{fig_en_vs_x1_x2}a, 
temperature independent since we are neglecting the electron entropy.  
In addition, we suppose that the displacements of nearest-neighbour 
sites are coupled to each other by an $SO(2)\cong U(1)$ invariant term 
that tends to minimise the strain. With those assumptions the classical 
Hamiltonian reads: 
\begin{equation}
  \label{Hph}
  \begin{split}
    \HH_\text{ph}( {\bf X} ) &=  
    J\sum_{\langle ij\rangle}\,
    \big( {\bf X}_i -{\bf X}_j\big)\cdot \big({\bf X}_i- {\bf X}_j\big) \cr
  & \phantom{+}+ \sum_i\, \Phi_\text{eff}\big({\bf X}_i\big)\,,
\end{split}
\end{equation}
where $\mathbf{X}$ denotes a configuration of all the displacement vectors. 
The model \eqn{Hph} is equivalent to a generalized $XY$-model, where 
${\bf X}_{i}$ plays the role of two-component spin of variable length, 
while $J>0$ is the conventional spin stiffness.
$\Phi_\text{eff}\big({\bf X}_i\big)$ is the effective anisotropic
potential obtained from the solution of the electron problem. 
Both the length and the direction of the local distortion  
${\bf X}_{i}$ are controlled by the effective potential 
$\Phi_\text{eff}\big({\bf X}_i\big)$, which is not invariant under $U(1)$  
but under separate $X_1\to -X_1$ and $X_2\to-X_2$ transformations, i.e., 
$Z_2 \times Z_2$. The phase diagram of an 
$XY$ model in presence of an anisotropy term that lowers $U(1)$ down to 
$Z_n$ is already known \cite{Schneider-PRL1976,Nelson-PRB1977,PhysRevLett.99.207203}. 
In particular, the anisotropy $Z_n$ for $n\geq 4$ is a dangerously irrelevant 
perturbation that does not change the $XY$ universality class of the transition 
\cite{Nelson-PRB1977,PhysRevLett.99.207203}. Our specific case study, 
where $U(1)\to Z_2\times Z_2$, has not been considered yet, at least to our 
knowledge, but it should most likely change the $XY$ universality class, which 
is what we are going to investigate in the following.      


We study the classical model \equ{Hph} at different temperatures using 
standard Monte Carlo (MC) method~\cite{Binder-MC}.
We consider the model on a three-dimensional 
cubic lattice of side $N_{x}$. 
The average value of a given observable, i.e. $\bra O \ket =
\frac{1}{Z}\sum_{\bf X} \OO({\bf  X})e^{-\beta H_{ph}({\bf X})}$,
where $Z=\sum_{\bf X} e^{-\beta  H_{ph}({\bf X})}$, is therefore 
estimated statistically using  MC algorithm to explore the
configuration space. New configurations are generated and
accepted/rejected using Metropolis algorithm with local updates. Each
local update corresponds to a shift $\Delta X_i$ of one of the two component
$i=1,2$, chosen with equal probability. Within our calculations we use
$\Delta X_i=0.15$, yet we checked that smaller values do not change
the accuracy of the calculations. 
In addition, we include the possibility of global moves of the type
$\bX_i \rightarrow \big(-X_{1i},X_{2i}\big)$, $\big(X_{1i},-X_{2i}\big)$ or $\big(-X_{1i},-X_{2i}\big)$ 
with a total probability $P_\textrm{global}=0.05$ equally
distributed among the three cases, i.e. with  probability
$P_\textrm{global}/3$ each. 
The local updates require the evaluation of the effective potential
$\Phi_\text{eff}\big({\bf X}_i\big)$ at the new value of
$\bX_i$. To speed-up execution, we pre-evaluate all the interpolated
values of the effective potential at any possible point compatible with the size
of the shift using a bi-cubic spline method. 
A new configuration of the system is obtained after a full
sweep of the lattice sites. 
The statistical error is thus controlled by the number of sweeps
$N_{s}$, to which it corresponds a number $N_s N_x^3$ of MC steps.
In order to avoid self-correlation problems, we measure the average of
any observables every $N_\textrm{meas}\simeq100$ sweeps and in any
case after a warm-up period of $N_\textrm{wp}\simeq1000$
sweeps. In all our calculations, the number of sweeps is of the
order of $N_s=4-6 \times 10^5$.
We further minimize the statistical error by executing the numerical
computation in parallel with $N_\textrm{cpu}=20$ cpu. The resulting
statistical error is within the symbols in all our plots.

Before discussing the results, we have to mention that some details
might depend on the precise form of the coupling between different
sites. In the model Hamiltonian \equ{Hph} we have choosen the simplest
possible one, i.e. a nearest neighbor coupling, thus disregarding
longer range interactions.

\begin{figure}
 \centering \includegraphics[width=0.475\textwidth]{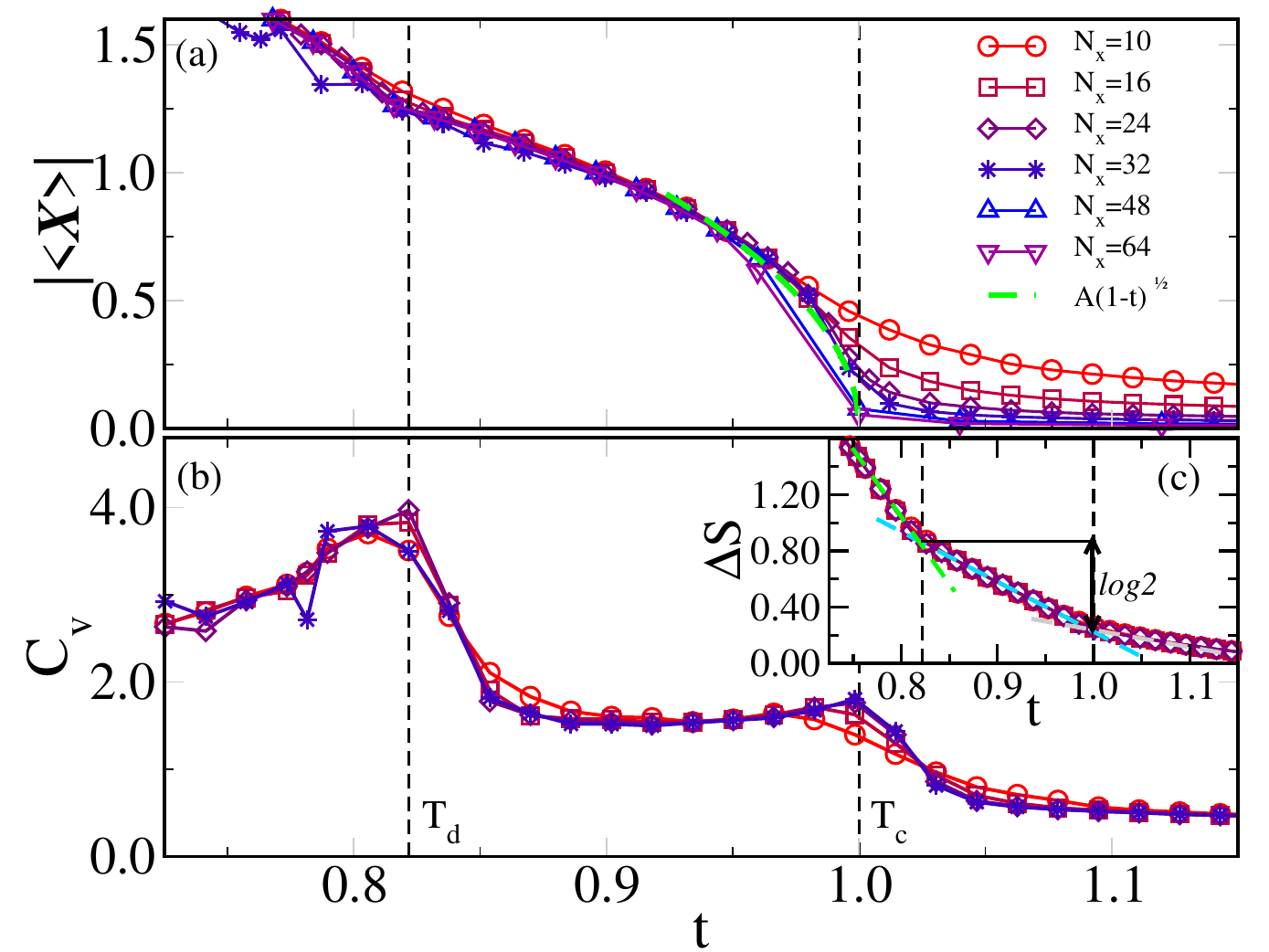}
  \caption{ (Color online) (a) Modulus of the average displacement as function of the
    reduced temperature $t=T/T_c$.
    (b) Specific heat for the same model as in panel (a) 
    as function of the reduced temperature. The data are for
    $N_s=4\times 10^5$ 
    MC sweeps of the lattice and
    different linear size $N_x$ (solid lines and open symbols). The
    dashed line in the critical region is a fit of the form
    $A(1-t)^{1/2}$, with $A=0.96$.
    (c) Entropy loss $\Delta S=S_\infty-S(t)$ as a function of reduced
    temperature and $N_x$ as in (a).  $S_\infty$ is the value attained
    in the limit of infinite temperature. The
    dashed lines are guides to the eye, emphasising the different
    linear behaviours of the entropy across the two phase transitions.
    }
  \label{fig_phonon1}
\end{figure}
\begin{figure*}
 \centering \includegraphics[width=1\textwidth]{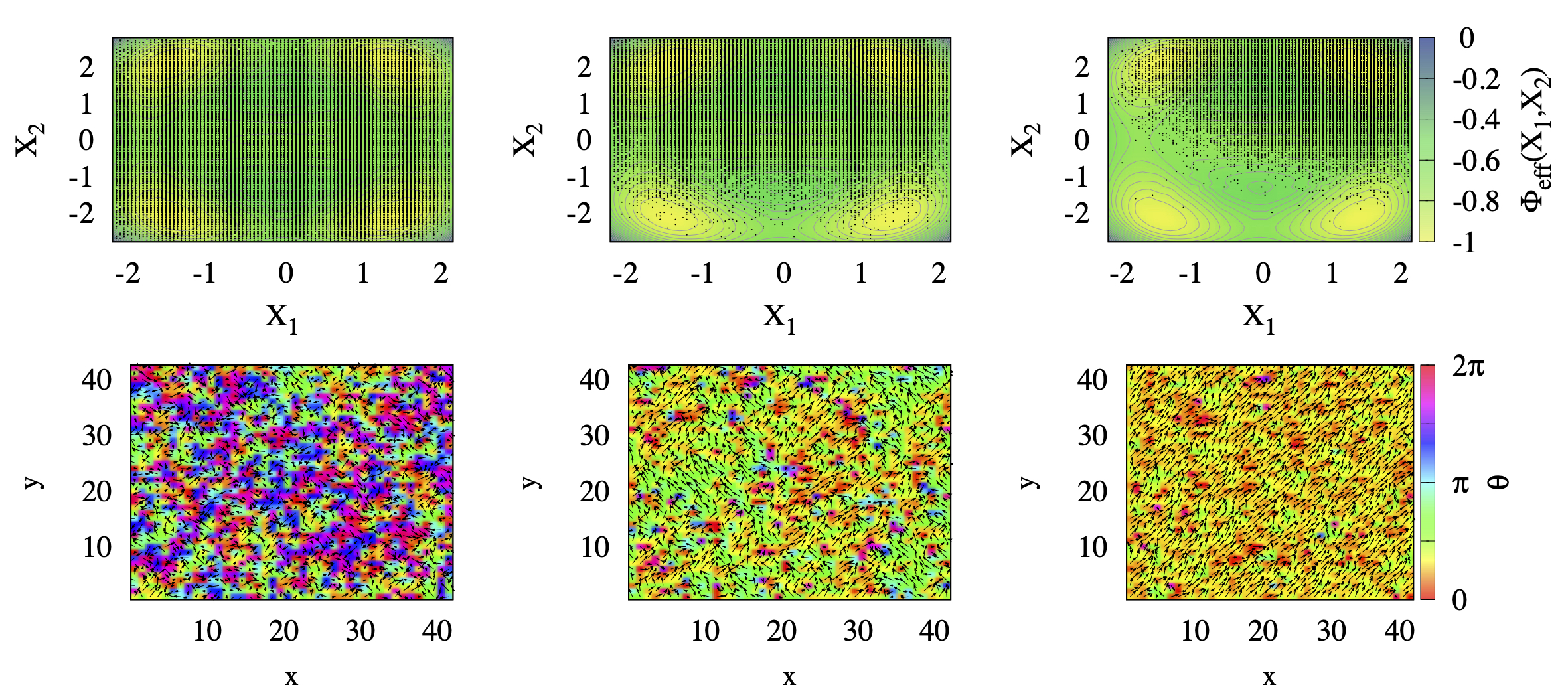}
 \caption{ (Color online)
   {\it Top panels}: distribution of the displacements $\bX_i$ at the
   end of the MC simulation, superimposed to the adiabatic potential
   $\Phi_\text{eff}$, properly normalised so that
   $\Phi_\text{eff}\in[-1,0]$.  
    Data are for $N_x=50$, $N_s=4\times 10^5$ MC sweeps of the lattice, and reduced temperatures
    $t=T/T_c$: $t=1.03$ (left), $0.82$ (center),  $0.71$ (right).
    Each (black) dot represents one of the $N_x^3$ endpoints of the calculation. 
    {\it Bottom panels}:  displacement field configuration within a single plane of
    the cubic lattice, with the same parameters of the top
    panels. If $\bX_i = \big|\bX_i\big|\,(\cos\theta_i,\sin\theta_i)$, 
    the color code represents $\theta_i\in[0,2\pi]$, and the arrow length 
    $\big|\bX_i\big|$. At high temperature $T>T_c$ (left panels)
    $\bX_i$ have random length and orientation, thus covering
    homogeneously the entire potential landscape.    
    For $T_d<T<T_c$ (center panels) the displacement orientation shows
    breaking of the $Z_2$ symmetry $X_2\to-X_2$.
    At lower temperature $T<T_d<T_c$, also the residual $Z_2$ symmetry
    $X_1\to-X_1$ gets broken; most of the $\bX_i$'s have length and
    direction corresponding to just one of the potential global
    minima.    
  }
  \label{fig_phonon3}
\end{figure*}
In \figu{fig_phonon1}(a) we plot the modulus of the
average displacement, $\big|\langle\,\bX\,\rangle\big|$, as function
of the temperature. 
For small system size (e.g. $N_x=10$)
$\big|\langle\,\bX\,\rangle\big|$ shows a smooth crossover in
temperature. However, increasing $N_x$ unveils the existence of a
continuous phase-transition at a critical value $T_c$ of the
temperature. 
The actual value of $T_c$ depends by the size of the coupling constant $J$, that somehow acts as a unit of measure for the energy. More involved calculations are necessary for the evaluation of $J$ in Vanadium dioxide from first-principle calculations and we postpone them for a future work. For that reason, we have preferred to use $T_c$ as the unit of temperature in \figu{fig_phonon1} and in those that follow.
In order to better reveal the second order character of the
transition, we also show in \figu{fig_phonon1}(a) the fit with a
mean-field square-root behaviour. The fit is rather good, although we
known that close to the transition the actual critical behaviour must
deviate from mean-field.



A closer look to the temperature dependence of the order parameter 
uncovers a non-trivial two-step evolution, which is more evident 
in \figu{fig_phonon1}(b), where we show the specific heat
$C_v=\partial \bra E\ket/\partial T$ vs. $T$. 
Indeed, $C_v$ clearly displays two peaks that are suggestive of two distinct  
transitions. The first transition at $T=T_c$, below which 
$\big|\langle\,\bX\,\rangle\big|$ acquires a finite value, is followed by a
second one at lower $T=T_d<T_c$.

From the knowledge of the specific heat at constant volume $C_v \left( T \right)$, we can compute the change of the vibrational entropy in the region where the two transitions occur as $\Delta S \left( t \right)= \int^{\infty}_{t} \frac{C_{V} \left( T \right)}{T} d T $. This quantity is displayed in \figu{fig_phonon1}(c) and it shows an almost linear behavior in the whole explored temperature range. However, as shown by the dashed lines there, the slopes of the line is different in the three regions $T<T_{d}$, $T_{d} < T < T_{c}$ and $T>T_{c}$.

\begin{figure}
 \centering \includegraphics[width=0.475\textwidth]{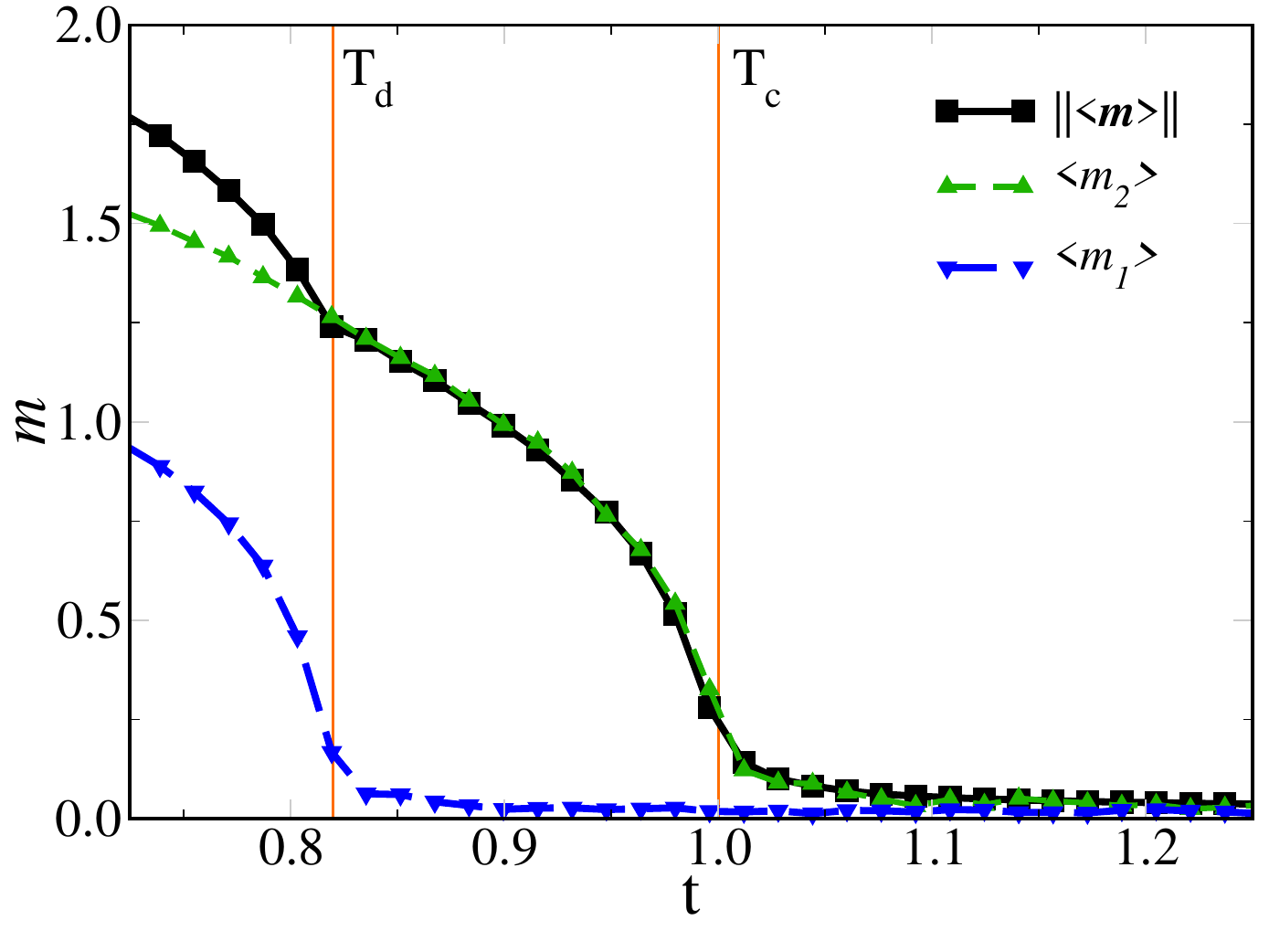}
 \caption{ (Color online) magnetization $m$ as a function of the
   reduced temperature $t=T/T_c$. Data are for $N_s=6\times 10^5$ MC
   sweeps, $N_x=24$.
   The solid line (black with open
   square) is the modulus of the average 
   magnetization vector. The dashed lines (green and blue with open
   triangles) indicate the behavior of the average of the
   magnetization components. The solid (orange) vertical lines
   indicate the two critical temperatures $T_d<T_c$ associated to the
   two stage transition. 
 }
  \label{fig_phonon4}
\end{figure}

In order to understand the nature of both transitions, in \figu{fig_phonon3} we 
show at $T>T_c$, left panels, $T_d<T<T_c$, middle panels, and $T<T_d$, 
right panels, the endpoint distribution after $N_s=4\times10^5$ MC sweeps 
of the lattice of the $N_x^3$ displacement vectors superimposed to the 
potential landscape in the $(X_1,X_2)$ space (top panels), and a real space 
snapshot within a single layer of the cubic lattice (bottom panels). At high 
temperature, $T>T_c$,  the $\bX_i$'s cover homogeneously the 
whole potential landscape, see top-left panel, without any appreciable spatial 
correlation, see the bottom-left panel. 
Lowering $T$ slightly below $T_c$, we observe a 
significant change in the displacement distribution, see middle panels. 
Specifically, the system seems to break ergodicity first along $X_2$, 
in the simulation corresponding to the figure it localises in the $X_2>0$ half-plane, 
while it is still uniform along $X_1$. Consequently, clusters 
of parallel displacement vectors 
form in real space. 
The alignment direction has $X_2>0$ for all clusters, 
while the $X_1$ component changes from cluster to cluster, see bottom-middle panel. 
Only below $T_d$, full ergodicity breakdown occurs, with the system 
trapped around just one of the four equivalent minima, in the figure that 
with $X_2>0$ and $X_1>0$. In other words, the $Z_2\times Z_2$ symmetry of the 
model \equ{Hph} gets broken in two steps upon cooling: first, the $Z_2$ symmetry 
$X_2\to-X_2$ spontaneously breaks, and next, the residual $X_1\to-X_1$ symmetry, 
leading to two consecutive Ising-like transitions. This is summarised in Fig.~\ref{fig_phonon4}, 
where we see that at $T_c$ $\langle X_2 \rangle$ becomes finite, and thus also 
$\big|\langle\bX\rangle\big|$, while $\langle X_1 \rangle$ is still zero. Only below 
$T_d$ also $X_1$ acquires a finite average value. 
Accordingly the vibrational entropy loss $\Delta S$, shown in
\figu{fig_phonon1}(c),  changes of  a quantity $\sim \ln(2)$ between
the two temperatures $T_{d}$ and $T_{c}$. This is consistent with an
increase of the available phase space for the system of a factor of
two by moving from one temperature to the other.

Translated in the language of VO$_2$, these results suggest the existence of an 
intermediate monoclinic phase for $T_d<T<T_c$ where the V atoms are displaced  
only within the basal plane, i.e., the chains are tilted but not yet dimerised. In our 
model Hamiltonian \eqn{hamiltonian}, such phase with $\langle X_1\rangle=0$ 
describes a monoclinic metal, which, as discussed in \secu{secI}, has been 
reported in several experiments 
\cite{doi:10.1143/JPSJ.21.1461,PhysRevB.77.235401,doi:10.1063/1.4764040,Yao-PRL2010,Laverok-PRL2014,doi:10.1021/nl9028973,PhysRevLett.97.266401,PhysRevB.85.155120,doi:10.1002/adma.201402404,Ilinskiy2012a,Ilinskiy2012b,doi:10.1063/1.3009569}. 
Only below $T_d$, both components of the antiferrodistortive displacement are 
finite, leading to the M1 insulating phase. 

In conclusion, without including the electron entropy we find two transitions 
that look continuous and in the Ising universality class: one at $T_d$ 
between a monoclinic insulator and a monoclinic metal, and another at 
$T_c>T_d$ from the monoclinic metal to a rutile one. On the contrary, 
neglecting the lattice entropy and just including the electronic one, we 
found in \secu{subsecIV.A} a single first-order transition at $T_\text{el}$, 
directly from the monoclinic insulator to the rutile metal. We can try now to 
argue what we could have obtained keeping both entropy contributions still 
within the Born-Oppenheimer adiabatic approximation. \\

In that case, we expect that the depth of the rutile minimum in the Born-Oppenheimer 
potential of Fig.~\ref{fig_en_vs_x1_x2} becomes a growing functions of $T$, unlike the depth of the 
insulating minima, since in the insulator the electronic entropy is negligible with respect to that in the metal. For the same reason, we expect that the height of the two equivalent saddle points at $X_1=0$ but 
$X_2\simeq \pm 2.5$, see black and blue lines in panel (b) and panel (c), respectively, of Fig.~\ref{fig_en_vs_x1_x2}, 
lowers with increasing $T$, since these points with large crystal field splitting but without dimerisation just describe the monoclinic metal, eventually turning these saddle points into local minima.  This effect might 
well turn the monoclinic-insulator to monoclinic-metal transition at $T_d$ into a first order one, all the more 
if we better modelled the martensitic features of the structural distortion. However, even in that case 
we still expect a further transition at $T_c>T_d$ into the rutile metal, unless the latter has such a large entropy compared to the monoclinic metal to drive a first order transition from the insulator directly 
into the rutile metal, as it would occur if $T_\text{el}< T_d$, i.e., if the electronic entropy gain far  
exceeds the lattice one.  \\
The experimental evidences supporting the existence of a monoclinic metal 
phase intruding between the M1 insulator and R metal 
\cite{doi:10.1143/JPSJ.21.1461,PhysRevB.77.235401,doi:10.1063/1.4764040,Yao-PRL2010,Laverok-PRL2014,doi:10.1021/nl9028973,PhysRevLett.97.266401,PhysRevB.85.155120,doi:10.1002/adma.201402404,Ilinskiy2012a,Ilinskiy2012b,doi:10.1063/1.3009569} 
suggest that, should our modelling be indeed representative of VO$_2$, then the 
Hamiltonian parameters should be such that $T_\text{el}\gtrsim T_d$. This also entails a substantial release of lattice entropy across the transition, in accordance with experiments \cite{Nature_ph,doi:10.1063/1.5042089} and theoretical \cite{PhysRevB.99.064113} proposals. We emphasise that 
$T_\text{el}\gtrsim T_d$ does not mean that correlations play a minor role, but rather the opposite, since 
it would imply the insulator, whose internal energy is substantially contributed by electronic correlations,  would survive up to much higher temperature if it were not for the lattice. 

\section{Conclusions}\label{secV}

We have constructed a minimal model that we believe contains all 
essential ingredients to correctly capture the physics of the 
metal-insulator transition in vanadium dioxide. 

The model comprises two orbitals per site, 
one mimicking the $\text{a}_{1g}$ and the other the $\text{e}^\pi_g$, 
thus neglecting the twofold nature of the latter, which broaden into 
two bands. 
The $\text{a}_{1g}$ band has a double peak structure reflecting its 
bonding character along the rutile $c$-axis, while the 
$\text{e}^\pi_g$ one is structureless. Both have the same bandwidth 
and centre of gravity. The density corresponds to one electron per 
site, i.e., the two bands are at quarter filling. The electrons feel 
an on-site Hubbard repulsion, and are coupled to 
two zone-boundary lattice modes, corresponding, respectively, to the 
basal plane component, i.e., the tilting of the Vanadium chains, and 
out-of-plane component, responsible of the chain dimerisation, of the 
antiferrodistortive displacement that acquires a finite expectation 
value below the transition from the high temperature rutile structure 
to the low temperature monoclinic one (M1). 
Using realistic Hamiltonian parameters and assuming the Born-Oppenheimer 
adiabatic approximation, we find at low temperatures phase coexistence 
between a stable distorted insulator, the monoclinic M1 insulator, and a 
metastable undistorted metal, the rutile metal. 
Upon rising temperature, our model description suggests a two-step transition. First, the 
dimerisation component of the antiferrodistortive displacement melts, 
leading to a transition from the monoclinic insulator to a monoclinic metal. 
At higher temperature also the tilting component disappears, and the 
monoclinic metal turns into the rutile one. Such a two-transition scenario, 
not in disagreement with experiments, is mostly driven by the lattice entropy, 
also in accordance with experiments. 

One of the messages of our model calculation is that the electron-electron 
interaction has the role to effectively enhance the coupling to the lattice, 
stabilising a distorted phase otherwise metastable in the absence of interaction. 
This also implies that we could have obtained similar results with 
weaker electronic correlations but stronger electron-lattice coupling. 
This conclusion is actually supported by the phenomenology of Niobium 
dioxide NbO$_{2}$, which, \textit{mutatis mutandis}, is akin to that of  VO$_2$. 
NbO$_{2}$ also undergoes a metal-insulator transition, though at 
substantially higher temperature of $T_\text{MIT} \sim 1080~\mathtt{K}$ 
\cite{JANNINCK19661183,Beebe:17,SETA1982921,RAO1973340}. There is some experimental evidence of separate structural and electronic phase transitions occurring in this compound \cite{SETA1982921,RAO1973340,GOODENOUGH1971145,PhysRevB.16.386}, with a transition temperature for the structural change $T_\text{s} \sim 1123~\mathtt{K}$ \cite{doi:10.1002/pssb.19670200263}, from a high-temperature rutile structure to a low-temperature body centred tetragonal (BCT) one that locally resembles the M1 phase of VO$_2$ \cite{GOODENOUGH1971145,doi:10.1063/1.4972830,BOLZAN19949,HIROI201547}. It has been proposed that the mismatch between $T_\text{MIT}$ and $T_\text{s}$ can be justified by invoking a melting of the dimerization component of the structural distortion in the BCT insulator at smaller temperature as compared to the melting of the tilting component \cite{GOODENOUGH1971145,RAO1973340,PhysRevB.16.386,doi:10.1143/JPSJ.26.582}. The metallic solution that appears in between the two transition temperatures should be mostly metallic along the c$_{R}$ axis, since the almost one dimensional $\text{a}_{1g}$ band gives the most relevant contribution to the spectral weight at the Fermi level in this intermediate phase. This expectation is not in disagreement with some experimental findings in which they measure, above $T_\text{MIT}$, a metallic conductivity along c$_{R}$ while a semiconducting one in the orthogonal direction \cite{doi:10.1139/p74-297}. However, we should point out that not all the experiments confirm this scenario \cite{PhysRevMaterials.3.074602}. We believe that a similar anisotropy in the conduction properties should be displayed also by the monoclinic metallic phase of Vanadium dioxide. 
The single $4d$-electron in Nb$^{4+}$ is expected to be less correlated 
than the $3d$-electron in V$^{4+}$. This loss of correlations, testified by the 
VO$_2$ M2 phase having no counterpart in NbO$_{2}$ \cite{PhysRevB.59.13650}, 
and by the efficacy of \textit{ab initio} methods to describe NbO$_{2}$ 
\cite{Eyert_2002,PhysRevB.96.195102,doi:10.1063/1.4903067,PhysRevB.91.094305}, 
is actually overcompensated by the increase in covalency due to the broader 
spatial distribution of the $4d$ orbitals \cite{PhysRevB.90.115135}, which, in turn, 
yields a stronger coupling with the zone-boundary lattice modes, and thus a 
higher transition temperature.

\section*{Acknowledgements}
F.G. likes to thank Sergiy Lysenko for the useful discussions about the 
thermodynamic potential for the two phononic modes, as well as for the 
choice of the parameters appearing there. F.G. thanks also Maja 
Berovi\'{c} and Daniele Guerci for the discussions concerning the 
manuscript. A.A. thanks Massimo Capone and Sandro Sorella for useful discussions.
We thank Martin Eckstein for fruitful debatings. 
We acknowledge support from the H2020 Framework 
Programme, under ERC Advanced Grant No. 692670 ``FIRSTORM''. 
A.A. also acknowledges financial support from MIUR PRIN 2015
(Prot. 2015C5SEJJ001) and SISSA/CNR project "Superconductivity,
Ferroelectricity and Magnetism in bad metals" (Prot. 232/2015).

\bibliography{biblio}{}
\end{document}